\documentclass[proof]{pasj00}

\begin{document}
\SetRunningHead{K.Yoshikawa et al.}{Non-equilibrium Ionization State}
\title{NON-EQUILIBRIUM IONIZATION STATE OF \\ WARM--HOT INTERGALACTIC MEDIUM}


\author{Kohji \textsc{Yoshikawa}}
\affil{Department of Physics, School of Science, University of Tokyo,
Tokyo 113-0033, Japan}
\email{kohji@utap.phys.s.u-tokyo.ac.jp}
\and
\author{Shin \textsc{Sasaki}}
\affil{Department of Physics, Tokyo Metropolitan University,
\\ 1-1 Minami-Osawa, Hachioji, Tokyo 192-0397, Japan}


\Received{}
\Accepted{}
\Published{}
\maketitle
\KeyWords{cosmology: miscellaneous --- X-rays: general --- methods: numerical}
\begin{abstract}
 Time evolution of the ionization state of metals in the cosmic baryons
 is investigated in a cosmological context without the assumption of
 ionization equilibrium. We find that a significant fraction of ionized
 oxygen (O\,{\sc vii} and O\,{\sc viii}) in the warm-hot intergalactic
 medium (WHIM) is not in the ionization equilibrium state at a redshift
 of $z\simeq0$. We also investigate the effect on the detectability and
 observables of WHIM as a consequence of such deviation from ionization
 equilibrium. It is found that the detectability of WHIM is not altered
 very much both through its emission and absorption signatures, but line
 ratios between O\,{\sc vii} and O\,{\sc viii} are significantly
 different from those in the ionization equilibrium state.
\end{abstract}

\section{INTRODUCTION}

After \citet{Fukugita1998} first pointed out the importance of studying
the cosmic baryon budget, it is now widely accepted that only less than
half of the cosmic baryons in the present universe have been detected,
and the remaining portion, ``the missing baryon'', has evaded the direct
detection so far. Numerical simulations by \citet{Cen1999} suggest that
approximately 30 to 50 \% of the cosmic baryons at $z=0$ are in the form
of the diffuse intergalactic medium with temperature of $10^5\,{\rm
K}<T<10^7\,{\rm K}$ which is called warm-hot intergalactic medium
(WHIM), and subsequent numerical simulations (e.g., \cite{Dave2001})
with different numerical schemes and resolutions also support this
picture consistently. Based on these numerical predictions, many
observational efforts have been made to reveal the existence of WHIM.

Since the observational signature of WHIM is very weak, the detection of
WHIM is observationally quite challenging and first proposed through its
metal absorption features in the bright background beacons such as QSOs
and blazers \citep{Hellsten1998, Perna1998, Fang2000, Cen2001,
Fang2001}. After the first detection of O\,{\sc vi} absorption lines in
the spectra of a bright QSO H1821+643 by \citet{Tripp2000} and
\citet{Tripp2001}, number of detections are reported through absorption
features of O\,{\sc vi}, O\,{\sc vii}, O\,{\sc viii} and Ne\,{\sc ix}
ions \citep{Nicastro2002, Fang2002, Mathur2003, Fujimoto2004}, but they
are rather tentative. Only recently, the detection with sufficient
signal-to-noise ratio is reported by \citet{Nicastro2005a} and
\citet{Nicastro2005b} in which absorption signatures of WHIM are found
at two redshift in the spectra of the blazer Mrk421 during its two
outburst phases. Future proposed missions such as {\it Constellation-X}
and {\it XEUS} are expected to detect numerous WHIM absorbers.
Detection of WHIM absorption in the spectra of afterglows of gamma-ray
bursts (GRBs) are also proposed by \citet{Elvis2003} and
\citet{Elvis2004} using a dedicated mission such as {\it Pharos}.
\citet{Kawahara2005} investigate the feasibility of such detections in a
realistic manner based on cosmological hydrodynamic simulations.  As an
alternative strategy, several tentative detections of WHIM through its
metal line emission are claimed by \citet{Kaastra2003}, and
\citet{Finoguenov2003} using {\it XMM-Newton} satellite. However, these
detections are not significant enough to exclude the possibility that
the observed emission lines are the Galactic ones because of the limited
energy resolution ($\simeq 80$eV) of the current X-ray
detectors. Recently, \citet{Yoshikawa2003}, \citet{Yoshikawa2004}, and
\citet{Fang2005} show that future X-ray missions equipped with a high
energy resolution spectrograph such as {\it DIOS} (Diffuse Intergalactic
Oxygen Surveyor) and {\it MBE} (Missing Baryon Explorer) can detect the
line emission of WHIM in a convincing manner. Future X-ray observations
of WHIM through its emission and absorption features are expected to
unveil the physical properties of WHIM beyond its existence.

So far, most of theoretical and observational studies on WHIM assume
that the cosmic baryons are in their ionization equilibrium under a
given physical condition, irrespective of their thermal
history. However, as shown in this paper, some portion of the cosmic
baryons is expected to deviate from ionization equilibrium depending on
its thermal history, since the timescales of ionization and
recombination processes can be comparable or even longer than the age of
the universe in a low density plasma such as WHIM.  In this paper, we
focus on the non-equilibrium ionization state of metals in WHIM. Since
ionized metals such as O\,{\sc vi--viii} and Ne\,{\sc ix} are supposed
to be good tracers of WHIM, the ionization state of such metals and
their deviation from the ionization equilibrium state are of great
importance in studying the detectability of WHIM and probing its physical
condition. By relaxing the assumption that the cosmic baryons are in the
ionization equilibrium, we follow the time evolution of the ionization
state of metals as well as hydrogen and helium in cosmological
hydrodynamic simulations, and investigate to what extent the effect of
the non-equilibrium ionization state affects the detectability and
observational features of WHIM.

The rest of the paper is organized as follows. The importance of
non-equilibrium treatment of ionization balance in WHIM is described in
section 2. In addition, technical details in computing time evolution of
ionization fractions of ions are also shown in section 2. In section 3,
we consider a simple model of thermal histories of the cosmic baryons,
and investigate the non-equilibrium ionization state of oxygen ions and
their deviations from the ionization equilibrium state to understand
general properties of the ionization history of oxygen in the cosmic
baryons.  The non-equilibrium ionization balance in the cosmological
hydrodynamic simulation is shown in section 4, and the effect of the
non-equilibrium ionization state on the detectability and observational
features of WHIM is investigated in section 5.  Finally, in section 6,
we summarize our conclusion and discuss other important physical
processes which may also affect the ionization state of WHIM. Throughout
this paper, unless otherwise stated, we adopt a spatially flat cosmology
with $\Omega_{\rm m}=0.3$, $\Omega_{\rm \Lambda}=0.7$, $\Omega_{\rm
b}=0.04$, and $h=0.7$, where $\Omega_{\rm m}$ is the density parameter,
$\Omega_{\Lambda}$ the dimensionless cosmological constant, $\Omega_{\rm
b}$ the baryon density parameter, and $h$ the hubble constant in units
of 100km\,s$^{-1}$\,Mpc$^{-1}$.

\begin{figure}[htbp]
 \leavevmode
 \begin{center}
  \FigureFile(120mm,65mm){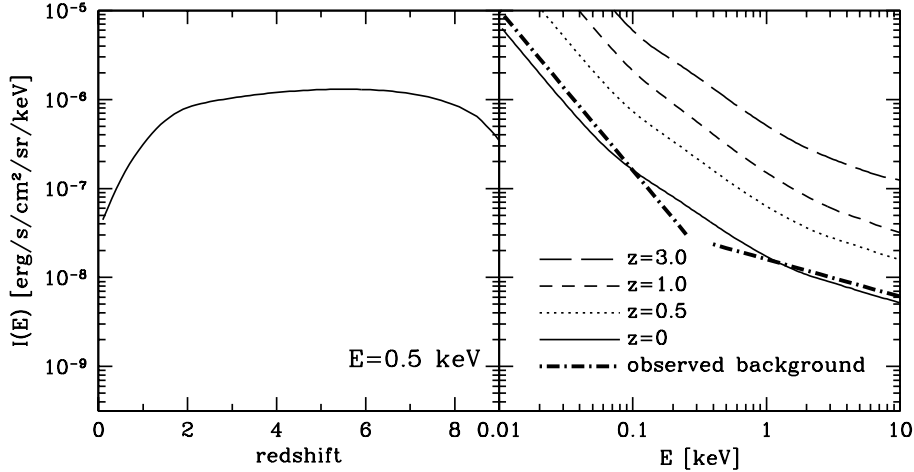}
 \end{center}
 \caption{Redshift dependence of the cosmic UV and X-ray background
 radiation at photon energy of 0.5 keV (left panel) and its spectrum
 at redshift $z=0$, 0.5, 1, and 3 (right panel). In the right panel,
 we also show the observed background spectra in UV and X-ray bands
 obtained by \citet{Shull1999} and \citet{Miyaji1998},
 respectively.}\label{fig:background}
\end{figure}

\section{IONIZATION BALANCE}

The time evolution of ionization balance in collisionally- and
photo-ionized plasmas can be described by
\begin{equation}
 \label{eq:evolution}
  \frac{df_j}{dt}=\sum_{k=1}^{j-1}S_{j-k,k}\,f_k-\sum_{i=j+1}^{Z+1}\,S_{i-j,j}\,f_j
  - \alpha_j\,f_j + \alpha_{j+1}\,f_{j+1},
\end{equation}
where $j$ is the index of a particular ionization stage considered, $Z$
the atomic number, $f_j$ the ionization fraction of an ion $j$. $S_{i,j}$
denotes the ionization rate for an ion $j$ with the ejection of $i$
electrons, and $\alpha_j$ is the recombination rate of ion $j$. As
ionization processes, collisional, Auger, charge-transfer and
photo-ionizations are taken into account. Recombination processes are
composed of radiative and dielectric recombinations. The ionization
equilibrium state, $f^{\rm eq}_j$, can be obtained by solving
\begin{equation}
 \label{eq:equilibrium}
  0=\sum_{k=1}^{j-1}S_{j-k,k}\,f^{\rm
  eq}_k-\sum_{i=j+1}^{Z+1}\,S_{i-j,j}\,f^{\rm eq}_j
  - \alpha_j\,f^{\rm eq}_j + \alpha_{j+1}\,f^{\rm eq}_{j+1}.
\end{equation}
Ionization and recombination rates adopted in this paper are calculated
by utilizing the SPEX\footnote{http://www.sron.nl/divisions/hea/spex/}
ver 1.10 software package.

\begin{figure}[htbp]
 \leavevmode
 \begin{center}
  \FigureFile(120mm,120mm){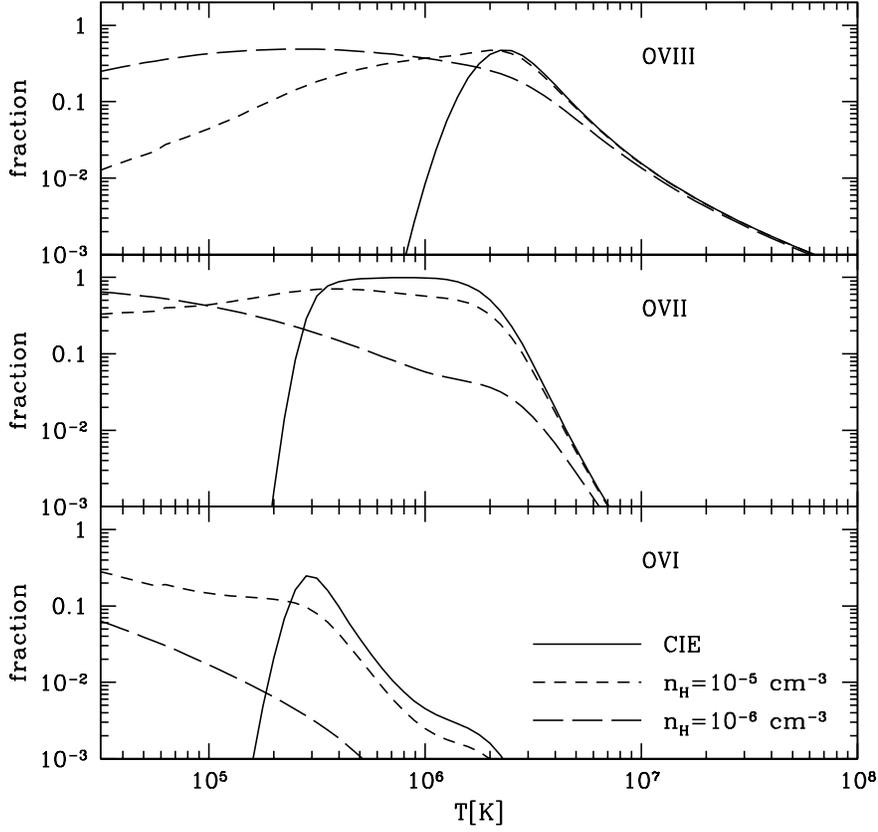}
 \end{center}
 \caption{Ionization fractions of O\,{\sc vi}, O\,{\sc vii} and O\,{\sc
 viii} in ionization equilibrium as a function of temperature for the
 low density plasma with $n_{\rm H}=10^{-5}$ and $10^{-6}$ cm$^{-3}$
 under the presence of photoionizing background radiation at a redshift
 of $z=0$. Ionization fractions in collisional ionization equilibrium
 are also shown for the comparison.\label{fig:fraction}}
\end{figure}

As a source of photoionization, we adopt UV and X-ray background
radiation calculated using the CUBA
code\footnote{http://pitto.mib.infn.it/\~haardt/cosmology.html}
\citep{Haardt2001} throughout this paper. The spectra and redshift
evolution of the adopted background radiation are shown in
figure~\ref{fig:background}. It should be noted that the intensity of
the background radiation rapidly drops at redshift $z<2$, while the
overall spectral shapes do not change so much. In the right panel, we
overlay the observed UV and X-ray background radiation estimated by
\citet{Shull1999} and \citet{Miyaji1998}, respectively, for
comparison. It can be seen that the spectrum of adopted background
radiation at $z=0$ is consistent with the observed ones within a factor
of two. Although ionization fractions of ions in the collisional
ionization equilibrium entirely depend on gas temperature, they depend
on gas density as well as gas temperature under the presence of
photoionizing background radiation, and the effect of photoionization by
the UV and X-ray background is important in a low density and low
temperature plasma. Figure~\ref{fig:fraction} shows the ionization
fractions of O\,{\sc vi}, O\,{\sc vii} and O\,{\sc viii} ions in
collisional ionization equilibrium (i.e. no background radiation) and
those in ionization equilibrium under the presence of the UV and X-ray
background at $z=0$. It is obvious that the effect of photoionization is
quite important at a typical temperature range of WHIM ($10^5\,{\rm K} <
T < 10^7\,{\rm K}$) and that it increases the ionization fractions of
ionized oxygen in a lower density regime.

\begin{figure}[thbp]
 \begin{center}
  \leavevmode
  \FigureFile(120mm,120mm){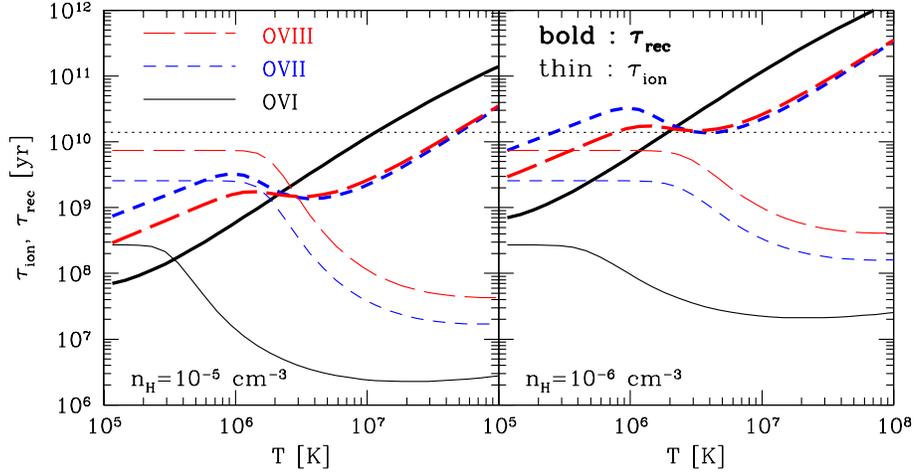}
 \end{center}
 \caption{Timescales of ionization and recombination for O\,{\sc vi},
 O\,{\sc vii} and O\,{\sc viii} ions in hydrogen number density of
 $n_{\rm H}=10^{-5} \mbox{cm$^{-3}$}$ (left) and $10^{-6}
 \mbox{cm$^{-3}$}$ (right) under the presence of UV and X-ray background
 radiation at a redshift of $z=0$. Horizontal dotted lines indicate the
 hubble time $H^{-1}_0$ ($\simeq 13$\,Gyr). \label{fig:timescale}}
\end{figure}

According to equation~(\ref{eq:evolution}), the timescales of ionization
and recombination for an ion $j$, $\tau_{\rm ion}$ and $\tau_{\rm rec}$
respectively, can be estimated by
\begin{equation}
 \tau_{\rm ion} = \left(\sum_{i=j+1}^{Z+1}S_{i-j,j}\right)^{-1}
\end{equation}
and
\begin{equation}
 \tau_{\rm rec} = \alpha_j^{-1}.
\end{equation}
Figure~\ref{fig:timescale} shows temperature dependence of $\tau_{\rm
ion}$ and $\tau_{\rm rec}$ of O\,{\sc vi}, O\,{\sc vii} and O\,{\sc
viii} ions for hydrogen number density of $n_{\rm H}=10^{-5}$ and
$10^{-6}$ cm$^{-3}$ under the presence of the photoionizing background
radiation at a redshift of $z=0$. In a temperature range
$T\gtrsim10^6$~K, the recombination timescales are longer than the
ionization timescales, and are comparable to or even longer than the
hubble time, $H_0^{-1}$ ($\simeq 13$\,Gyr), where $H_0=100h$
km\,s$^{-1}$\,Mpc$^{-1}$ is the hubble constant. In a lower temperature
regime with $T\lesssim10^6$ K, the recombination timescales are still
comparable to the hubble time for hydrogen number density of $n_{\rm
H}=10^{-6}$ cm$^{-3}$. The ionization timescales in the lowest
temperature range ($\simeq 10^5$ K) are determined by the intensity of
the UV and X-ray background radiation. It can be seen that the
ionization timescales of O\,{\sc vii} and O\,{\sc viii} ions are also
comparable to the hubble time at a temperature of $T\lesssim10^6$ K.
Actually, these behaviors of ionization and recombination timescales are
also the case for hydrogen- and helium-like ions of other metal such as
carbon and nitrogen. Considering these facts, one can easily expect that
the ionization balance is not in the equilibrium state in a low density
plasma like WHIM. Therefore, for proper treatment of ionization balance
in WHIM, we have to follow the time evolution of ionization fractions by
directly integrating equation~(\ref{eq:evolution}).

Since the timescales for the ionization and recombination differ by many
order of magnitude depending on ionic species and ionization stages,
equation~(\ref{eq:evolution}) is a stiff set of differential
equations. In numerically solving a stiff set of equations, explicit
integration schemes are unstable unless unreasonably small timestep is
taken, which leads to prohibitively expensive computations. Therefore,
implicit schemes are required to numerically integrate equation
(\ref{eq:evolution}). In this paper, we adopt a backward difference
formula (BDF), which is also adopted in \citet{Anninos1997} for solving
chemical reactions of atomic and molecular hydrogen in the primordial
universe. In the BDF, ionization fractions of all ionic species are
sequentially updated in the order of increasing ionization states rather
than being updated simultaneously, and the source terms contributed by
the ionization from and recombination to the lower states are evaluated
at the advanced timestep. The discretization of equation
(\ref{eq:evolution}) can be written as
\begin{equation}
 f_j^{t+\Delta t} = \frac{\left(\sum_{k=1}^{j-1}S_{j-k,k}f_k^{t+\Delta
			   t}+\alpha_{j+1}f_{j+1}^{t}\right)\Delta t+f^{t}_j}
 {1+\left(\sum_{i=j+1}^{Z+1}S_{i-j,j}+\alpha_j\right)\Delta t}.
 \label{eq:BDF}
\end{equation}
We confirm that this scheme is stable and as accurate as the eigenvalue
decomposition scheme by \citet{Kaastra1993} adopted for solving the
non-equilibrium ionization state in supernova remnants, but our scheme
is found to be computationally less expensive. In this paper, we
simultaneously solve the time evolution of ionization fractions of H,
He, C, N, O, Ne, Mg, Si, and Fe, but we focus on only the ionization
states of oxygen.

Throughout this paper, we assume spatially uniform and time independent
metallicity of $Z=0.1Z_{\odot}$, where $Z_{\odot}$ denotes the solar
abundance. Note that the metallicity $Z$ is very insensitive to the
resulting non-equilibrium ionization state as long as $Z\ll 1$. Of
course, in reality, metals in intergalactic medium are ejected from
galaxies through the feedback of star formation activities, and thus the
distribution of metals is inhomogeneous and time dependent. Actually,
for a fully consistent treatment, such feedback effect must be included
in the rhs of equation~(\ref{eq:evolution}) as a source term. In this
paper, however, since we are interested in the non-equilibrium
ionization state that arise along the thermal histories of baryons in a
context of cosmological structure formation, we adopt such an assumption
and believe that it is acceptable for qualitative studies of ionization
states of WHIM.

\section{SIMPLE MODELS}\label{sec:simple_model}

\begin{figure}[htbp]
 \begin{center}
  \leavevmode
  \FigureFile(120mm,120mm){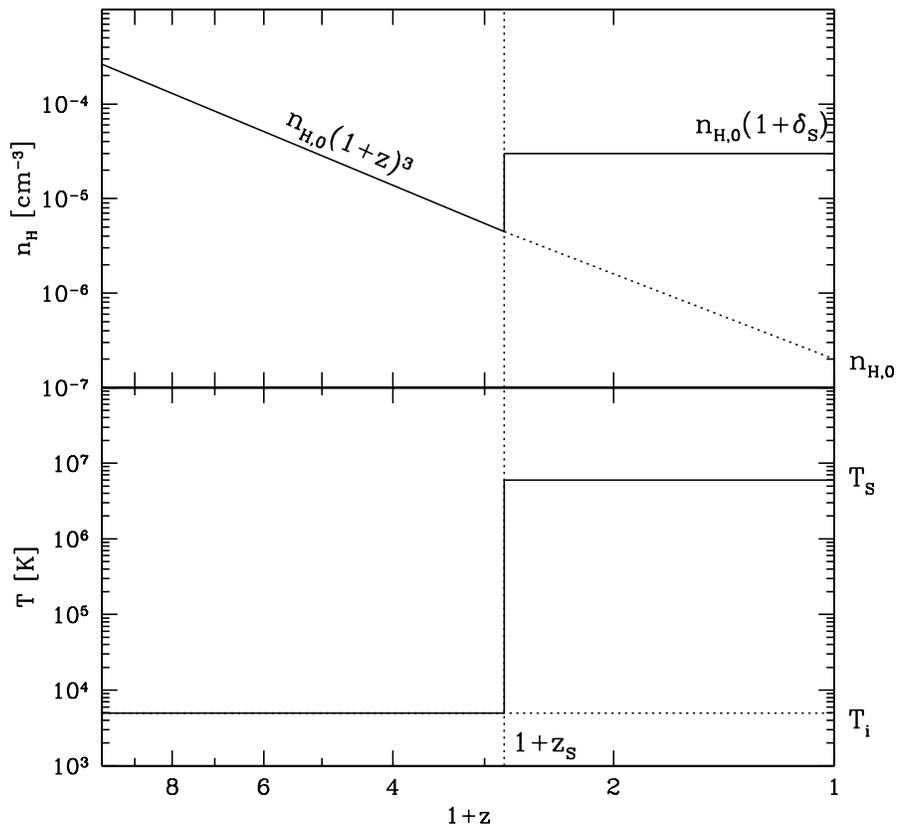}
 \end{center}
 \caption{Time evolution of density and temperature of baryons for the
 model calculations. We have three free parameters $\delta_{\rm s}$,
 $T_{\rm s}$ and $z_{\rm s}$. (see text for detail.)\label{fig:simple_model}}
\end{figure}

In this section, we explore the general behaviors of the non-equilibrium
ionization state using simple thermal histories of the cosmic
baryons. Here, we consider a series of thermal histories depicted in
figure~\ref{fig:simple_model}, in which baryonic matter has hydrogen
number density $n_{\rm H,0}(1+z_i)^3$ and temperature $T_i$ at an
initial redshift of $z_i=10$, keeps its comoving density and temperature
constant before increasing its density and temperature to $n_{\rm
H}=n_{\rm H,0}(1+\delta_{\rm s})$ and $T=T_{\rm s}$, respectively, by
experiencing shock heating at redshift $z=z_{\rm s}$, and keeps the
postshock thermal state till $z=0$. A physical scenario we have in mind
is that the baryonic matter is diluted according to the cosmological
expansion until it experiences the cosmological shock heating, and that
it gets quasi-virialized inside collapsed halos or filamentary
structures after the shock heating. Here $n_{\rm H,0}$ is the
present-day hydrogen number density, and we set the initial temperature
to $T_i=5\times10^3\,{\rm K}$. Therefore, these thermal histories can be
parametrized by three parameters, the postshock overdensity $\delta_{\rm
s}$, postshock temperature $T_{\rm s}$, and the redshift of shock
heating $z_{\rm s}$.  In a cosmological model adopted in this paper, we
have $n_{\rm H,0}=2.2\times10^{-7}$ cm$^{-3}$. We assume the UV and
X-ray background radiation described in the previous section as a
photoionization source, irrespective of the adopted thermal histories.

\begin{figure}[tbp]
 \leavevmode
 \begin{center}
  \FigureFile(150mm,150mm){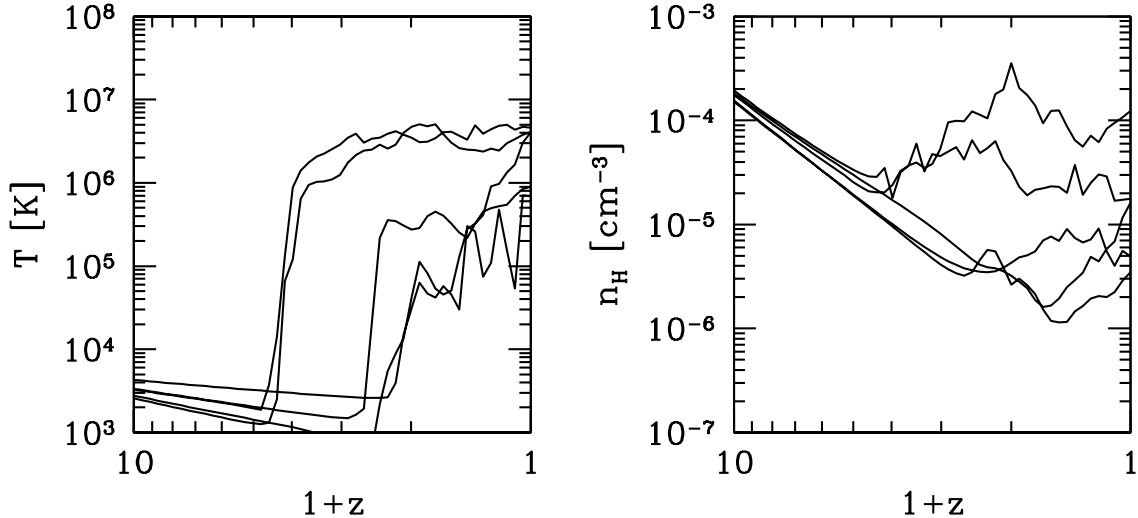}
 \end{center}
 \caption{Time evolution of temperature (left panel) and hydrogen number
 density (right panel) of gas particles in a cosmological hydrodynamic
 simulation presented in section \ref{sec:simulation}. See section
 \ref{sec:simulation} for details of the
 simulation. \label{fig:trajectory}}
\end{figure}

Actually, the thermal history of the cosmic baryon in a cosmological
hydrodynamic simulation is fairly well represented by this simple model.
Figure~\ref{fig:trajectory} shows the time evolution of temperature and
hydrogen number density of gas particles from a cosmological
hydrodynamic simulation presented in section~\ref{sec:simulation}.  In
both panels, time evolutions of five randomly selected gas particles
inside a filamentary structure at $z=0$ are shown. One can see that each
gas particle have nearly constant temperature and physical density
proportional to $(1+z)^{-3}$ before it experiences the first shock
heating. After the first shock heating, although thermal histories of
gas paricles in the simulation is not so simple as that presented in
figure~\ref{fig:simple_model}, it can be a viable model for qualitative
description of the thermal history of the cosmic baryon since the
timescales for temporal variations in density and temperature are much
shorter than $\tau_{\rm ion}$ and $\tau_{\rm rec}$ of oxygen ions
(O\,{\sc vi}, O\,{\sc vii} and O\,{\sc viii}). It should be noted that,
in figure~\ref{fig:trajectory}, we can see that some particles
experience multiple shocks in a sufficiently short timescale compared
with $\tau_{\rm ion}$ and $\tau_{\rm rec}$, and increase their postshock
density by a factor larger than 4, the value in the strong shock
limit. Therefore, we allow the density jump at $z=z_{\rm s}$ to be
greater than a factor of 4 in order to bracket the plausible range of
the thermal histories of the cosmic baryons.

At the initial redshift $z_i=10$, we assume that all baryons are in the
ionization equilibrium state. This assumption can be justified by the
fact that physical gas density is so high at such a high redshift that
ionization equilibrium can be quickly
realized. Equation~(\ref{eq:evolution}) is integrated along the thermal
histories described above till $z=0$ for a given set of parameters
$\delta_{\rm s}$, $T_{\rm s}$, and $z_{\rm s}$.
Figure~\ref{fig:contour_map_simple_zs_1.0} shows the ratios of the
resulting ionization fractions at $z=0$ for thermal histories with
$z_{\rm s}=1.0$ relative to those in the ionization equilibrium state,
$f_j/f_j^{\rm eq}$, for O\,{\sc vi}, O\,{\sc vii}, O\,{\sc viii} and
O\,{\sc ix} on a $(1+\delta_{\rm s})$--$T_{\rm s}$ plane. White contour
lines are separated by a factor of 1.1 in ratios of ionization
fractions, $f_j/f^{\rm eq}_j$, where the bold lines are contours for
$f_j/f_j^{\rm eq}=1$. Black contours with numeric labels indicate the
ionization fractions in the ionization equilibrium state on
$(1+\delta_{\rm s})-T$ plane. One can see that, for O\,{\sc vi} ion, the
ratios of ionization fractions significantly deviate from unity only in
the region where O\,{\sc vi} fraction in the equilibrium state, $f^{\rm
eq}_{\rm OVI}$, is negligible, and close to the equilibrium value in the
region where $f^{\rm eq}_{\rm OVI}$ is higher than 0.1. This is because,
as can be seen in figure~\ref{fig:timescale}, timescales for both of
ionization and recombination of O\,{\sc vi} ions are much shorter than
the hubble time in the density and temperature ranges where $f^{\rm
eq}_{\rm OVI}$ is significant. On the other hand, for O\,{\sc vii},
O\,{\sc viii} and O\,{\sc ix}, we have significant deviation from the
ionization equilibrium state in those density and temperature ranges in
which their ionization fractions are not negligible, especially at low
density regions with $1+\delta_{\rm s}<1.5$. Only the region with
$1+\delta_{\rm s}>10$ and $T_{\rm s}>10^{6.5} {\rm K}$ is almost free
from significant deviation from the ionization equilibrium state.

Figure~\ref{fig:trajectory_simple_zs_1.0} shows the time evolution of
ionization fractions of O\,{\sc vii}, O\,{\sc viii} and O\,{\sc ix} for
thermal histories with four sets of $1+\delta_{\rm s}$ and $T_{\rm s}$
marked in figure~\ref{fig:contour_map_simple_zs_1.0} (A, B, C and D) and
$z_{\rm s}=1.0$. For all the thermal histories, ionization fractions of
O\,{\sc viii} increase rapidly just after the shock heating, while those
in the ionization equilibrium state are lower than the preshock
ionization fractions. This is because the timescale for collisional
ionization to O\,{\sc viii} ion is much shorter than that of
recombination timescale. After taking their maximum values, O\,{\sc
viii} fractions decrease toward the equilibrium fractions. Ionization to
O\,{\sc ix} ions slightly slower and is delayed compared with that to
O\,{\sc viii} ions. Ionization fractions in the equilibrium state also
change in time because the intensity of the UV and X-ray background
radiation is decreasing at a redshift $z\lesssim2$.  In the history A
with $1+\delta_{\rm s}=10$ and $T=10^6$ K, O\,{\sc viii} and O\,{\sc ix}
fractions in the ionization equilibrium state are increasing and
decreasing in time, respectively. In the non-equilibrium ionization
state, however, the ionization fractions of these ions cannot catch up
with the ionization equilibrium state because of slow recombination
processes from O\,{\sc ix} to O\,{\sc viii}. Therefore, it can be said
that the decrease of the UV and X-ray background radiation at
$z\lesssim2$ is so rapid that it eventually enlarge the deviation from
the ionization equilibrium state. The similar behavior can be also seen
in the histories C and D. On the contrary, in the history B with higher
density and temperature than A, the ionization equilibrium is achieved
because ionization processes are dominated by collisional ionization and
the decrease in UV and X-ray background intensity does not significantly
change the equilibrium ionization
fractions. Figures~\ref{fig:contour_map_simple_zs_0.5} and
\ref{fig:contour_map_simple_zs_0.1} are the same as
figure~\ref{fig:contour_map_simple_zs_1.0} but for results with $z_{\rm
s}=0.5$ and $0.1$, respectively. As is the case for the results with
$z_{\rm s}=1.0$, we also have significant deviation from ionization
equilibrium for O\,{\sc vii}, O\,{\sc viii} and O\,{\sc ix} ions, while
O\,{\sc vi} fractions are close to the ionization equilibrium states for
the region where $f^{\rm eq}_{\rm OVI}>0.1$. For $z_{\rm s}=0.5$ and
$0.1$, fractions of O\,{\sc vii} and O\,{\sc viii} are larger than the
equilibrium fractions, while O\,{\sc ix} ion is under-ionized in regions
with $1+\delta_{\rm s}<10$ and $T_{\rm s}>10^6$K. These features can be
understood by the same argument as the result for a history with $z_{\rm
s}=1.0$ that the timescale for ionization to O\,{\sc viii} is much
shorter than that of recombination from O\,{\sc viii} to O\,{\sc vii},
and that ionization to O\,{\sc ix} is slower than that to O\,{\sc viii}.

\begin{figure}[htbp]
 \begin{center}
  \leavevmode
  \rotatebox{270}{\FigureFile(60mm,60mm){figure6a_color.eps}}
  \rotatebox{270}{\FigureFile(60mm,60mm){figure6b_color.eps}}
  \rotatebox{270}{\FigureFile(60mm,60mm){figure6c_color.eps}}
  \rotatebox{270}{\FigureFile(60mm,60mm){figure6d_color.eps}}
 \end{center}
 \caption{Maps of ratios of ionization fractions between non-equilibrium
 and equilibrium ionization states for O\,{\sc vi}, O\,{\sc vii},
 O\,{\sc viii} and O\,{\sc ix} at redshift $z=0$ on a $(1+\delta_{\rm
 s})$--$T_{\rm s}$ plane, where shock heating epoch is set to $z_{\rm
 s}=1.0$. White contours are separated by a factor of 1.1 in the ratios
 of the ionization fractions. Black lines with numeric labels are the
 contours of the ionization fractions in the ionization equilibrium
 states.\label{fig:contour_map_simple_zs_1.0}}
\end{figure}

\begin{figure}[htbp]
 \begin{center}
  \leavevmode
  \FigureFile(120mm,120mm){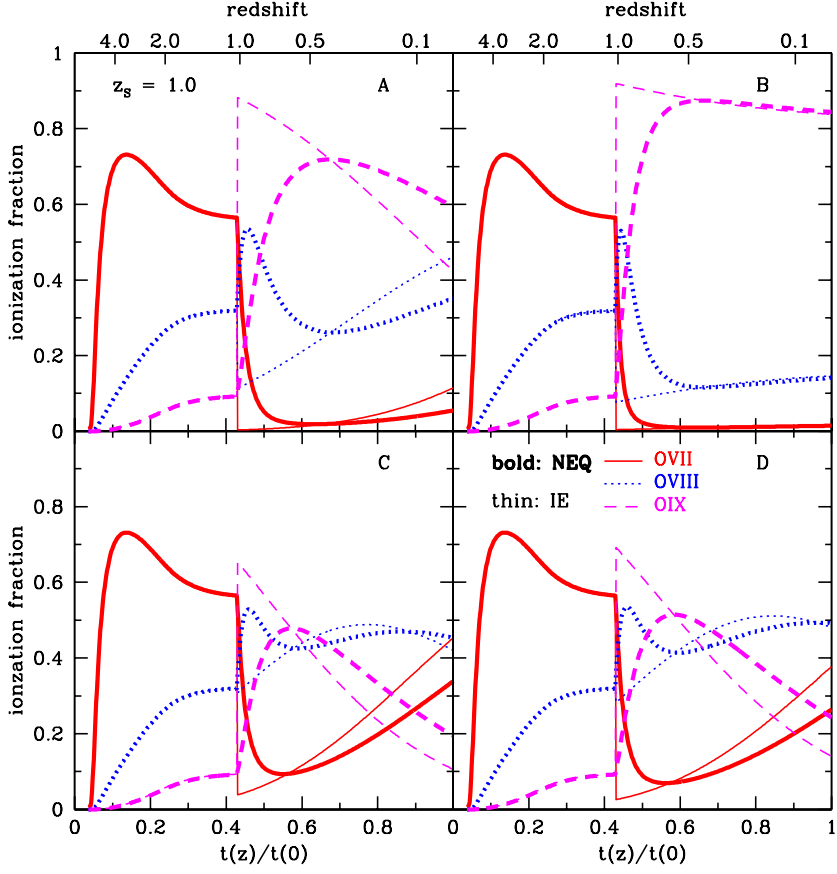}
 \end{center}
  \caption{Time evolution of ionization fractions of O\,{\sc vii},
  O\,{\sc viii} and O\,{\sc ix} for four sets of the densities and
  temperatures at redshift $z=0$ marked in
  figure~\ref{fig:contour_map_simple_zs_1.0}. Bold and thin lines
  indicate the non-equilibrium and equilibrium states, respectively.
  \label{fig:trajectory_simple_zs_1.0}}
\end{figure}

\begin{figure}[htbp]
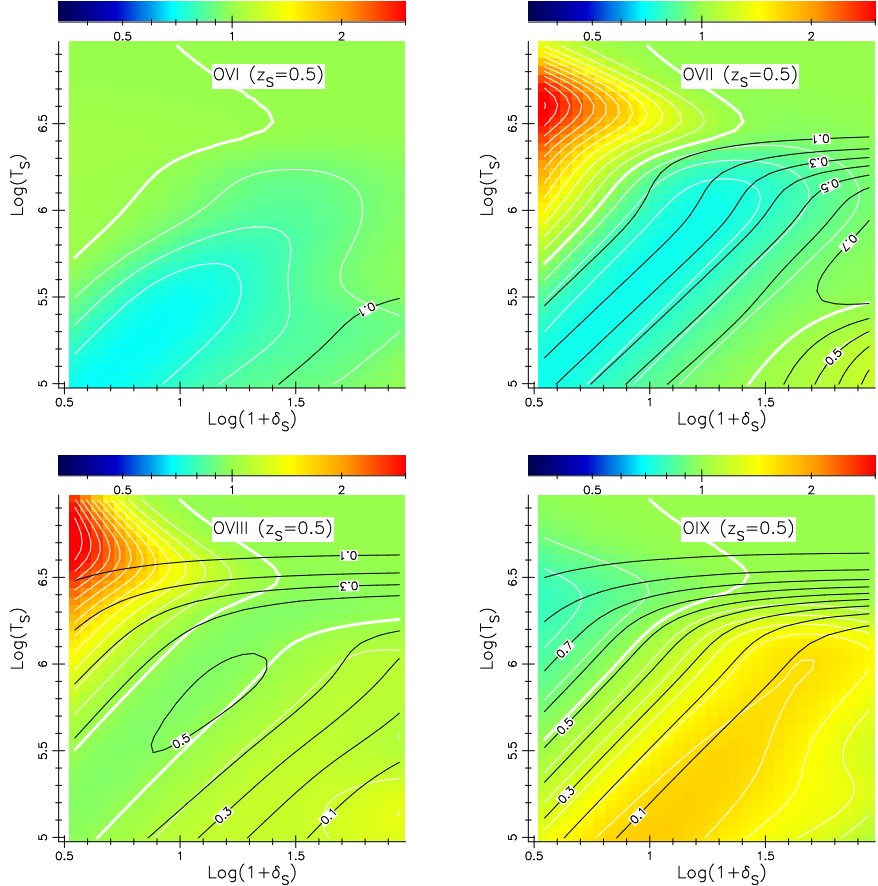

 \begin{center}
  \leavevmode
  \rotatebox{270}{\FigureFile(60mm,60mm){figure8a_color.eps}}
  \rotatebox{270}{\FigureFile(60mm,60mm){figure8b_color.eps}}
  \rotatebox{270}{\FigureFile(60mm,60mm){figure8c_color.eps}}
  \rotatebox{270}{\FigureFile(60mm,60mm){figure8d_color.eps}}
 \end{center}
 \caption{Same as figure~\ref{fig:contour_map_simple_zs_1.0} except that
 shock heating epoch is set to $z_{\rm
 s}=0.5$. \label{fig:contour_map_simple_zs_0.5}}
\end{figure}

\begin{figure}[htbp]
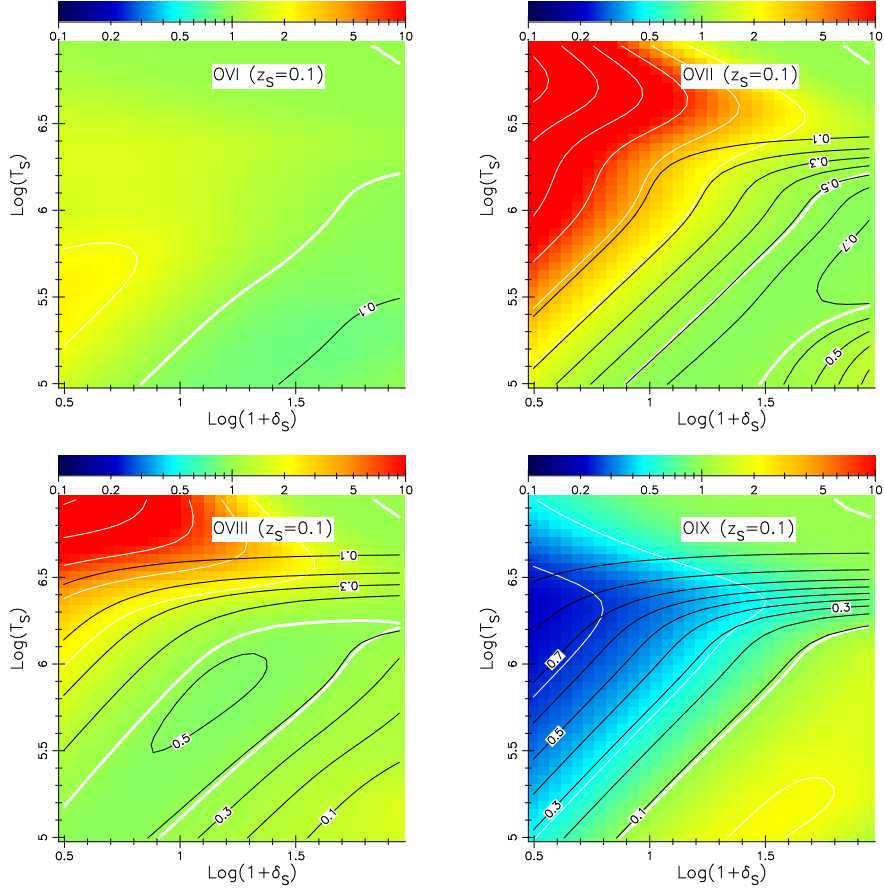

 \begin{center}
  \leavevmode
  \rotatebox{270}{\FigureFile(60mm,60mm){figure9a_color.eps}}
  \rotatebox{270}{\FigureFile(60mm,60mm){figure9b_color.eps}}
  \rotatebox{270}{\FigureFile(60mm,60mm){figure9c_color.eps}}
  \rotatebox{270}{\FigureFile(60mm,60mm){figure9d_color.eps}}
 \end{center}
 \caption{Same as figure~\ref{fig:contour_map_simple_zs_1.0} except that
 shock heating epoch is set to $z_{\rm s}=0.1$. Note that the ranges of
 the ratios are different from those in
 figure~\ref{fig:contour_map_simple_zs_1.0} and
 \ref{fig:contour_map_simple_zs_0.5}.
 \label{fig:contour_map_simple_zs_0.1}}
\end{figure}

\section{RESULTS IN COSMOLOGICAL SIMULATION}\label{sec:simulation}

In this section, we compute the non-equilibrium ionization state for
more realistic thermal histories of baryons obtained from a cosmological
hydrodynamic simulation. The simulation data of \citet{Yoshikawa2003} is
used in this paper, which is computed using a hybrid code of a
Particle--Particle--Particle--Mesh (PPPM) poisson solver and smoothed
particle hydrodynamics (SPH) scheme. The details of our code are
described in \citet{Yoshikawa2001}. We adopt a spatially flat
$\Lambda$CDM universe with $\Omega_{\rm m}=0.3$, $\Omega_{\Lambda}=0.7$,
$\Omega_{\rm b}=0.015h^{-2}$, $\sigma_8=1.0$ and $h=0.7$, where
$\sigma_8$ is the r.m.s. density fluctuation smoothed over a scale of
$8h^{-1}$Mpc. We employ $128^3$ particles each for dark matter and gas
particles within a periodic simulation box of $75h^{-1}$Mpc per
side. Therefore, the mass of each dark matter and gas particle is
$2.2\times10^{10}M_{\odot}$ and $2.4\times10^9M_{\odot}$,
respectively. The effect of radiative cooling is incorporated using the
cooling rate for a metallicity of [Fe/H]$=-0.5$ by
\citet{Sutherland1993}. The effect of energy feedback from supernovae is
ignored in our simulation.

The initial condition is setup at $z=36$ and we have 51 outputs from
$z=9$ to $z=0$ with an equal interval in $\ln(1+z)$. For each gas
particle, we integrate equation~(\ref{eq:evolution}) till $z=0$ using
the BDF described by equation~(\ref{eq:BDF}) using its thermal history,
or more specifically, a history of its density and temperature. As an
intial condition of equation~(\ref{eq:evolution}), the output at $z=9$ is
adopted. As in the simple model, it is assumed that all baryons are in
the ionization equilibrium state in the initial epoch. Thermal history
of each gas particle is interpolated if an time interval between two
continuous outputs is longer than the timestep $\Delta t$ in integrating
equation~(\ref{eq:BDF}). The calculations are carried out with timesteps
of $10^7$ yr and $3\times10^6$ yr for a convergence test, and we verify
that the results with two different timesteps are consistent with each
other. Note that our methodology is not fully self-consistent because
the thermal history of each gas particle, with which its non-equilibrium
ionization state is calculated, is computed with the cooling rate in the
collisional ionization equilibrium by \citet{Sutherland1993}. However,
since the cooling time for hot ($T\simeq 10^6$K) and low density
($1+\delta_{\rm b}\simeq 1$) gas like WHIM is much longer than the
hubble time, the effect of radiative cooling on its thermal history is
rather mild, and such inconsistency in radiative cooling does not
drastically change its thermal history. Therefore, we believe that our
results presented in this paper is useful as a first approximation of
non-equilibrium ionization state in WHIM.

\begin{figure}[tbp]
 \begin{center}
  \leavevmode
  \rotatebox{270}{\FigureFile(60mm,60mm){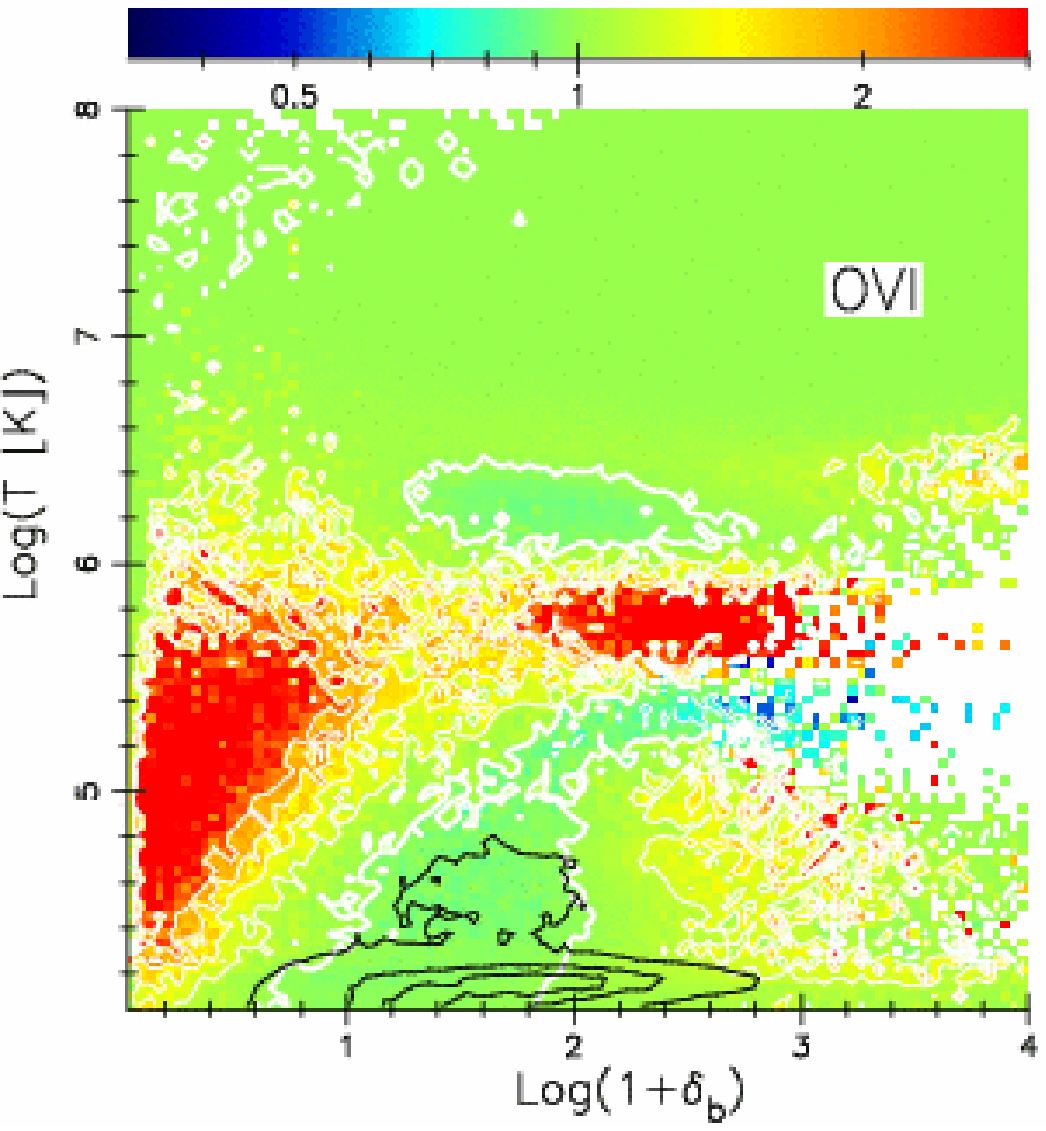}}
  \rotatebox{270}{\FigureFile(60mm,60mm){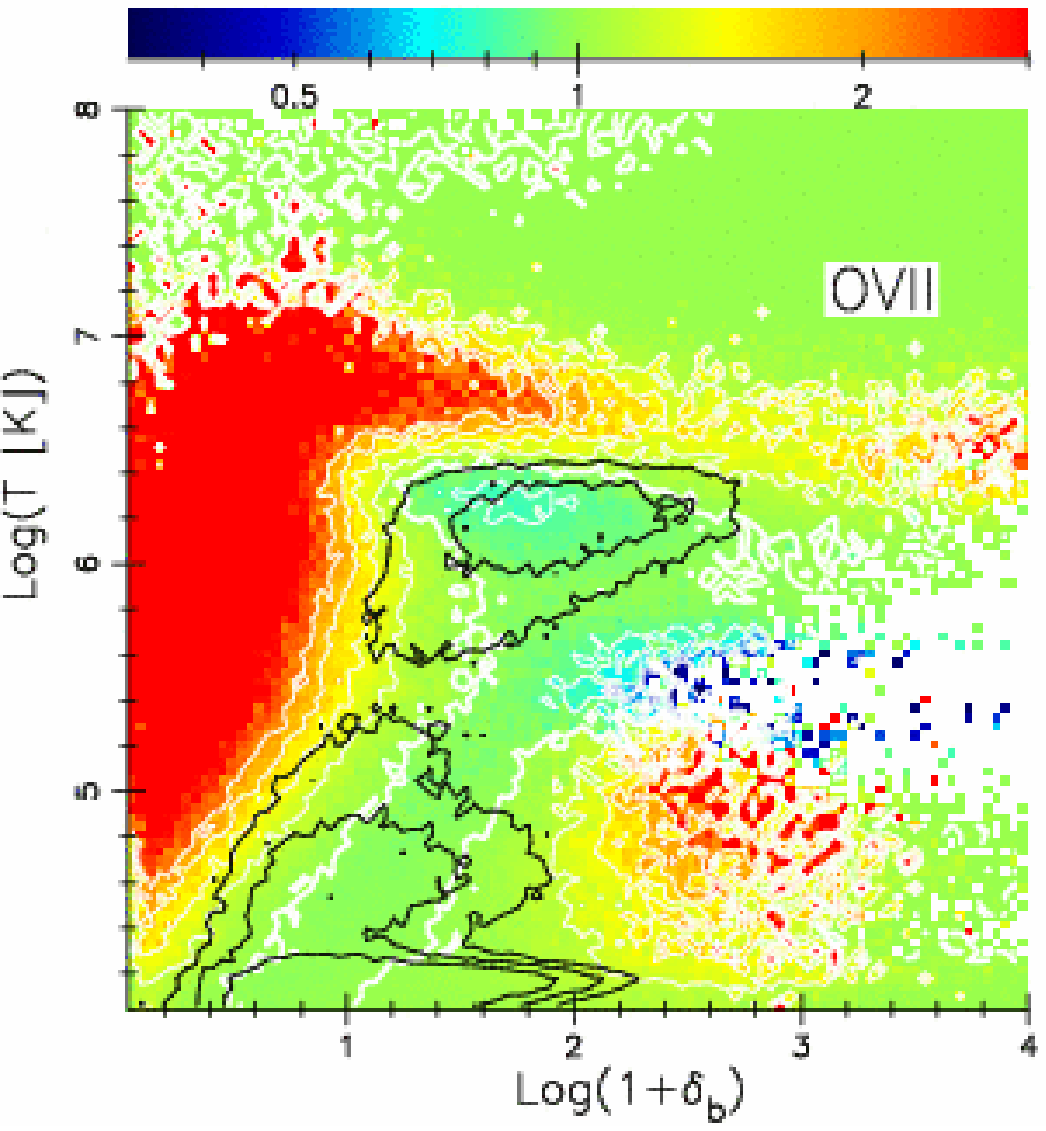}}
  \rotatebox{270}{\FigureFile(60mm,60mm){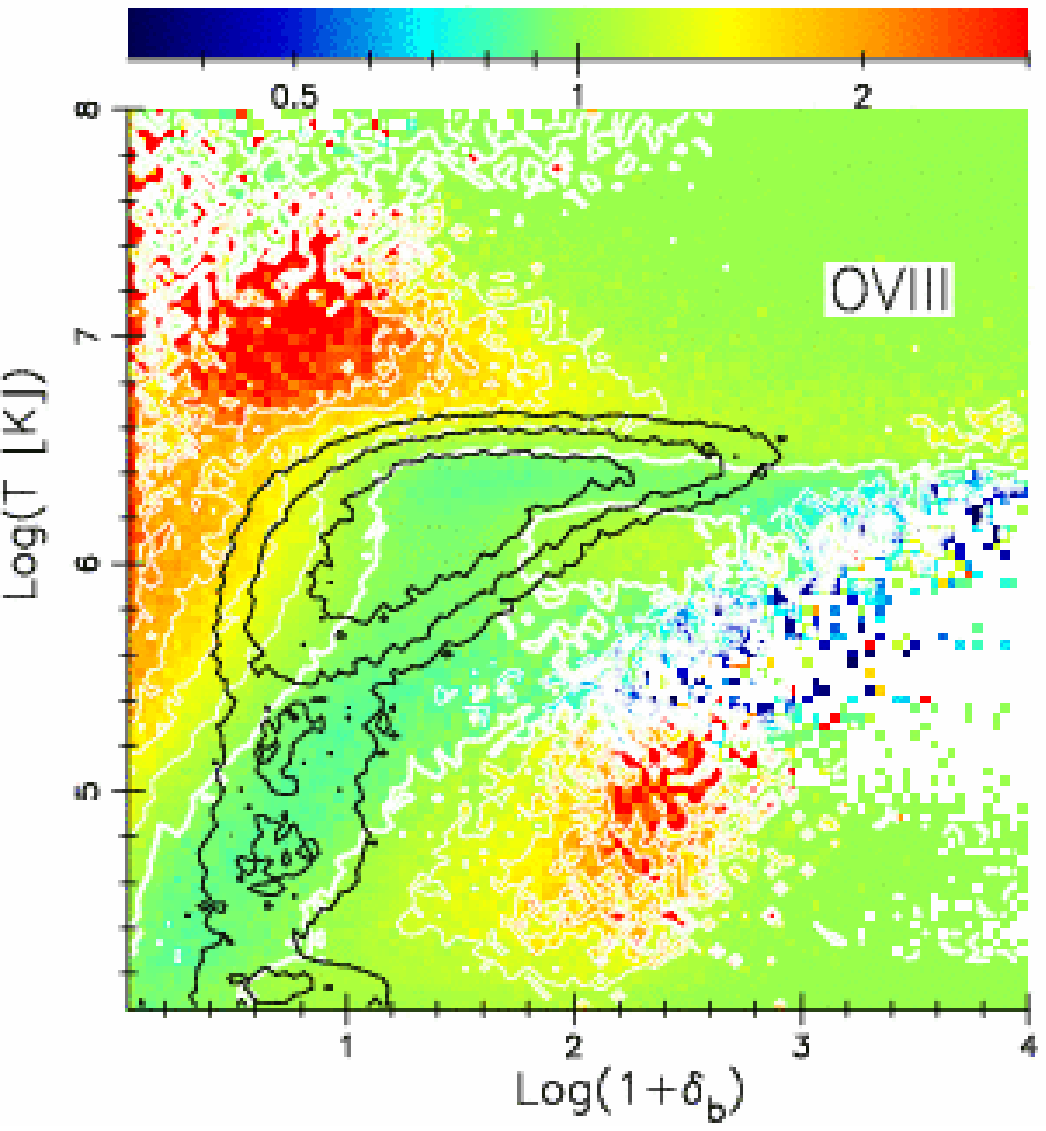}}
  \rotatebox{270}{\FigureFile(60mm,60mm){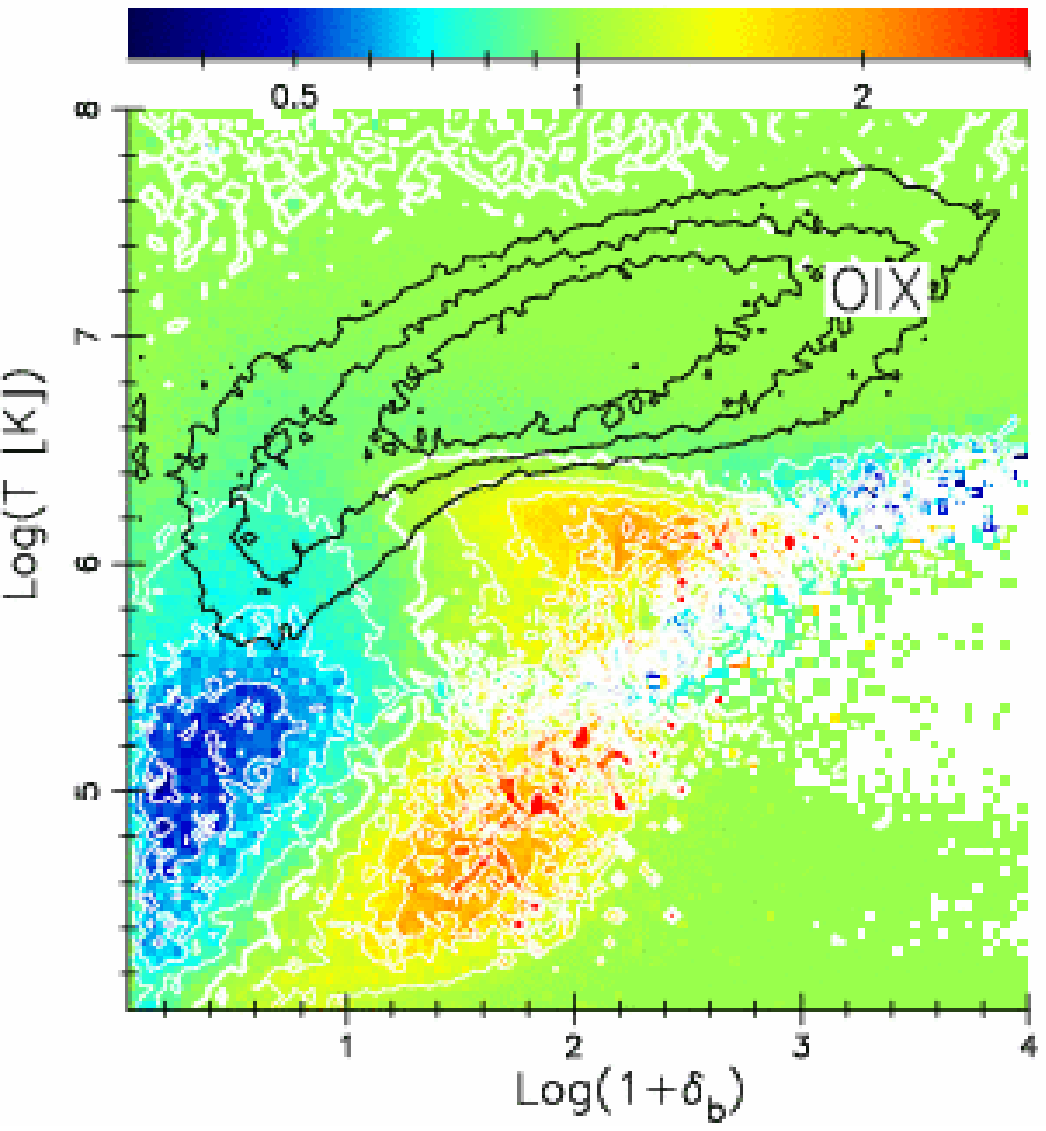}}
 \end{center}
 \caption{Mean ratios of ionization fractions at $z=0$
 between non-equilibrium and equilibrium ionization states for O\,{\sc
 vi}, O\,{\sc vii}, O\,{\sc viii} and O\,{\sc ix} on the
 $(1+\delta_{\rm b})$--$T$ plane. White contours are drawn in the regions with
 $0.5<f_j/f^{\rm eq}_j<2$ and separated by a factor of 1.2 in the
 ratios of ionization fractions. The white solid lines indicate the
 contour for $f_j/f^{\rm eq}_j=1$. Three black contours from inside to
 outside enclose the 25\%, 50\% and 75\% of mass of each ionic species.
 \label{fig:contour_map_simulation}}
\end{figure}

Figure~\ref{fig:contour_map_simulation} depicts mean ratios of
ionization fractions of O\,{\sc vi}, O\,{\sc vii}, O\,{\sc viii} and
O\,{\sc ix} in the non-equilibrium state at $z=0$ relative to those in
the equilibrium state on a $(1+\delta_{\rm b})$--$T$ plane, where
$\delta_{\rm b}$ is the baryon overdensity. Black contours in each panel
indicate the mass distribution of the corresponding ion, and 25\%, 50\%
and 75\% of its mass are enclosed by three contour lines from inside to
outside. One can see that most of O\,{\sc vi} ions are close to their
ionization equilibrium states, and that, the deviations of O\,{\sc vii},
O\,{\sc viii} and O\,{\sc ix} ions from the ionization equilibrium are
not negligible. In addition, all of O\,{\sc vii}, O\,{\sc viii} and
O\,{\sc ix} ions with $(1+\delta_{\rm b})>10^{2}$ and $T>10^{6.5}$K are
close to the ionization equilibrium. These features are similar to the
results for simple thermal histories described in the previous section,
and can be understood by the same arguments on the timescales of
ionization and recombination processes. It should be also noted that
deviation from ionization equilibrium states at low density regimes with
$1+\delta_{\rm b}\lesssim10$ is quite similar to that in the simple
models with $z_{\rm s}=0.1$ (see
figure~\ref{fig:contour_map_simple_zs_0.1}), indicating that the baryons
in these low density regions recently experienced shock heating.

\begin{figure}[htbp]
 \begin{center}
  \leavevmode
  \rotatebox{270}{\FigureFile(60mm,60mm){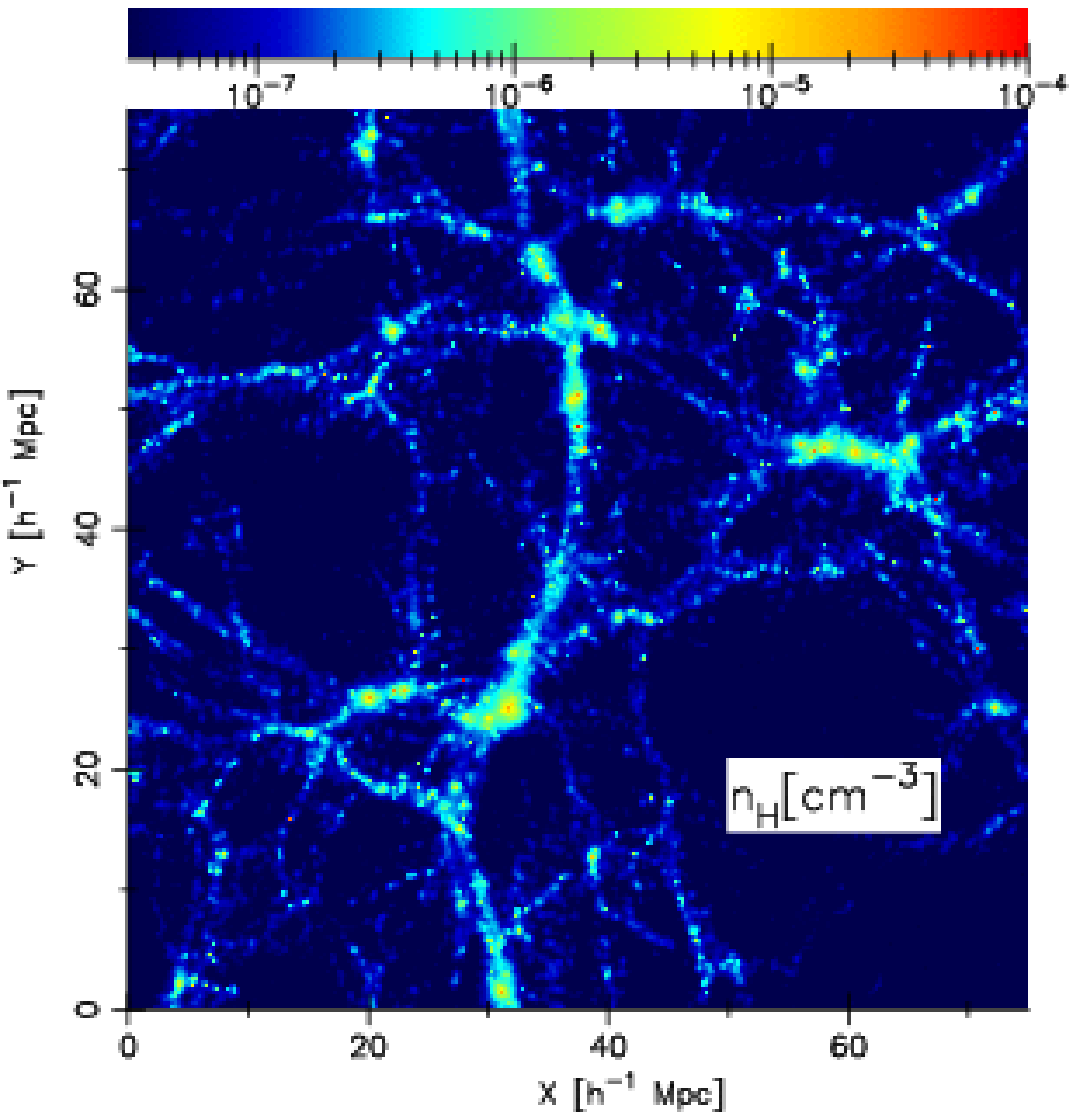}}
  \rotatebox{270}{\FigureFile(60mm,60mm){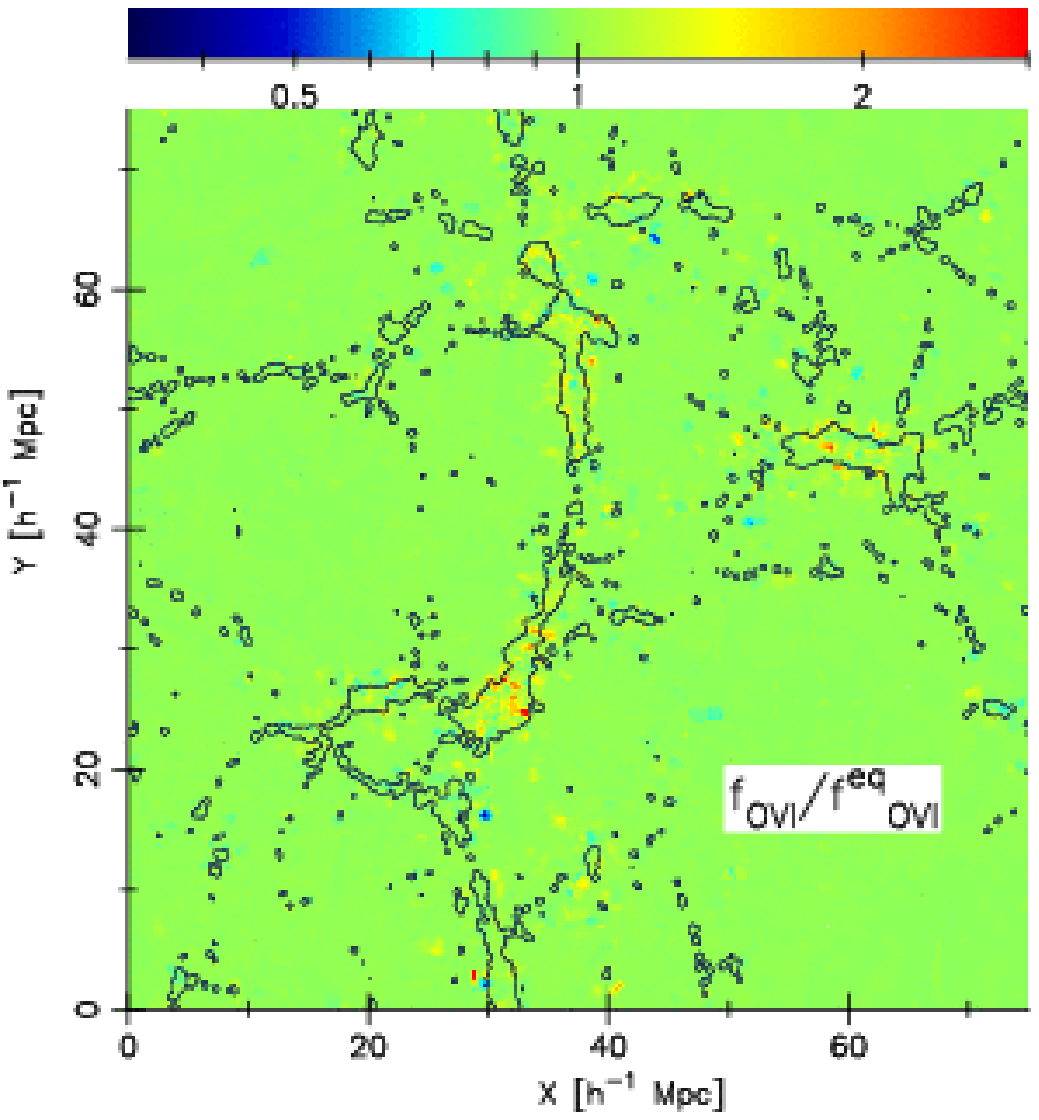}}
  \rotatebox{270}{\FigureFile(60mm,60mm){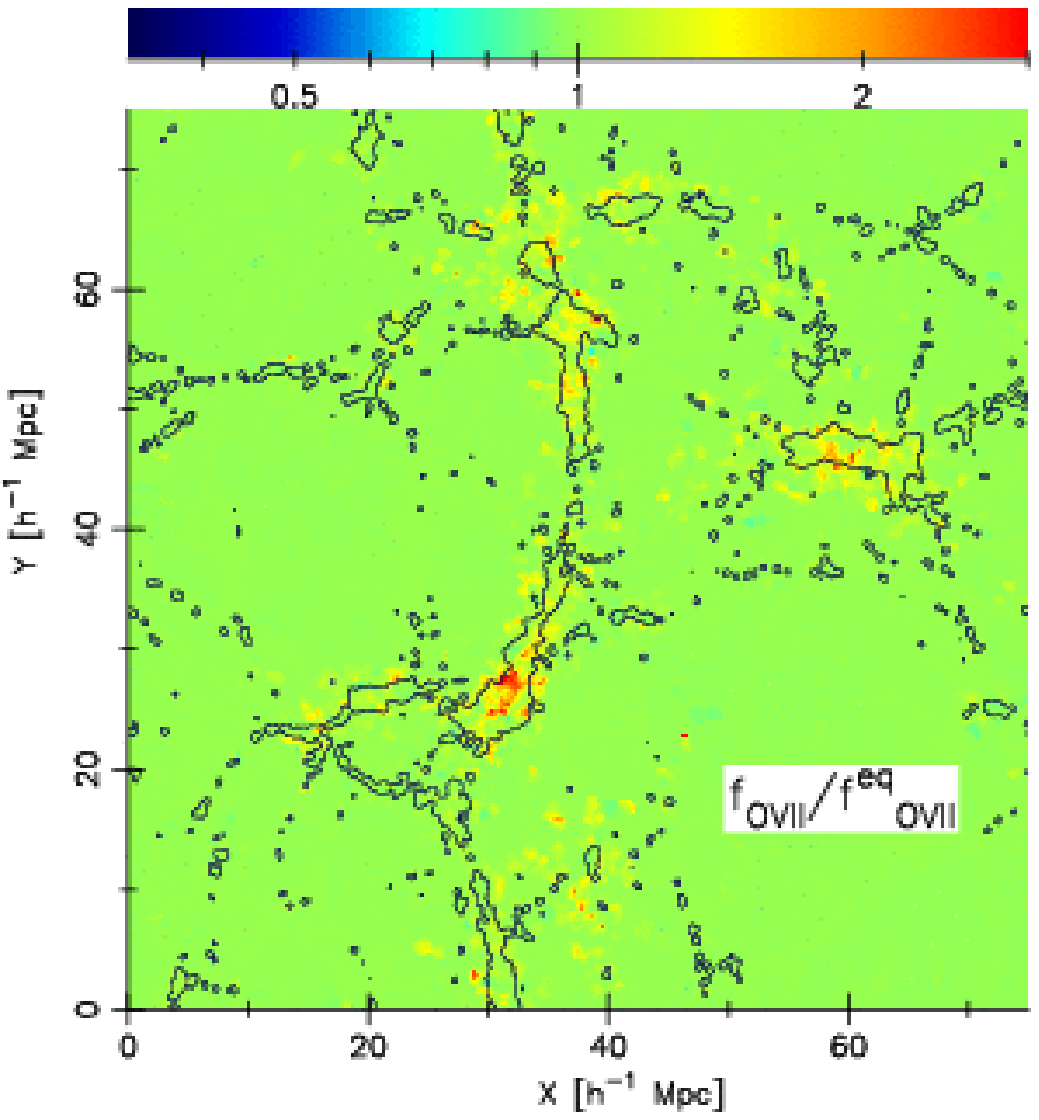}}
  \rotatebox{270}{\FigureFile(60mm,60mm){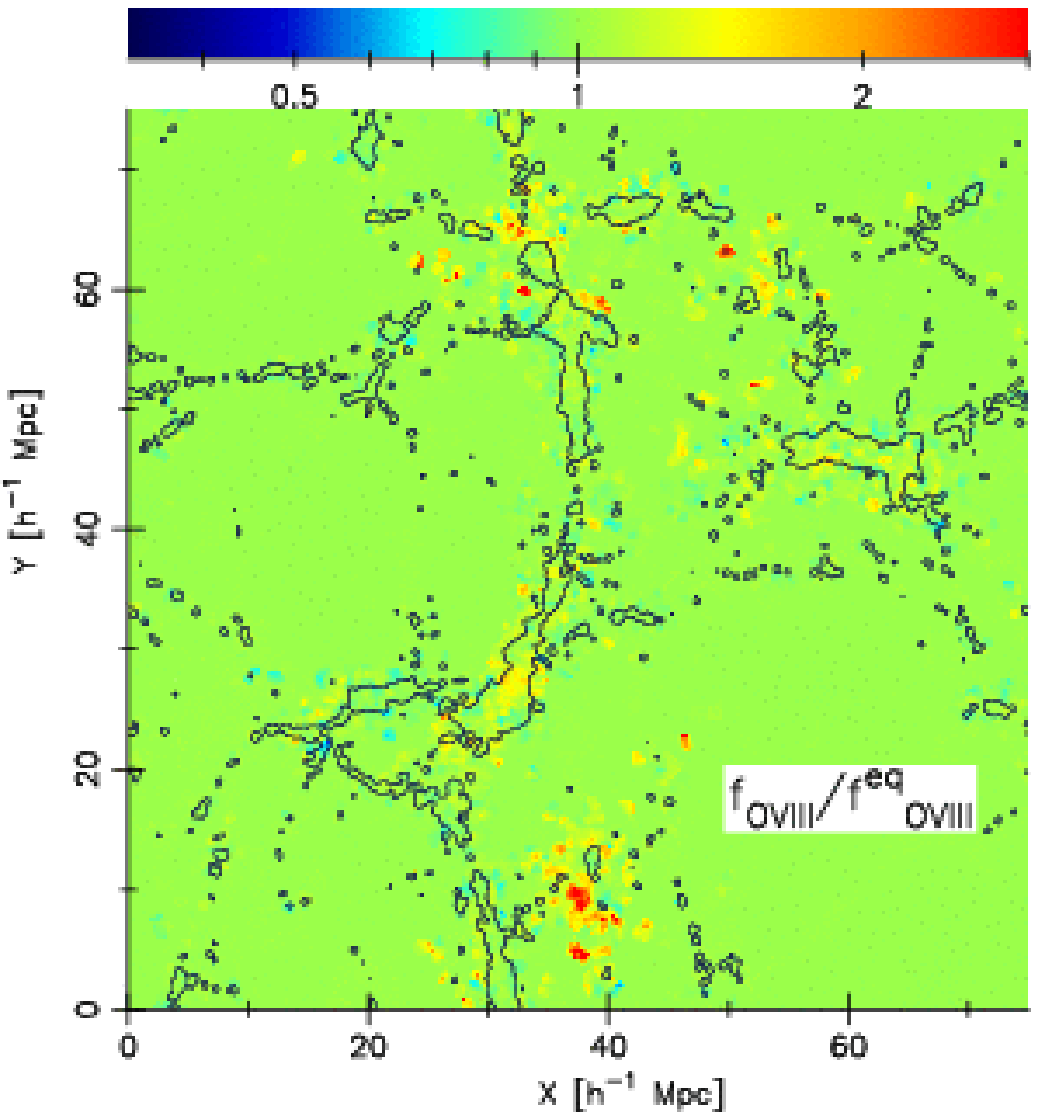}}
 \end{center}
 \caption{Spatial distribution of hydrogen number density, and ratios of
 ionization fractions of O\,{\sc vi}, O\,{\sc vii} and O\,{\sc viii}
 between non-equilibrium and equilibrium states in the simulation volume
 at $z=0$. The thickness of the slice is $7.5h^{-1}$Mpc. Contours of
 baryon mass density for $1+\delta_{\rm b}=1$ are also shown in the
 top-right, bottom-left and bottom right
 panels. \label{fig:distribution}}
\end{figure}

In figure~\ref{fig:distribution}, we show the spatial distribution of
hydrogen number density and the ratios of O\,{\sc vi}, O\,{\sc vii} and
O\,{\sc viii} fractions in the non-equilibrium state relative to those
in the ionization equilibrium state in a $7.5h^{-1}$ thick slice taken
from the simulation box at $z=0$. In this slice, we have a galaxy
cluster at $(X,Y)=(32,25)h^{-1}\,{\rm Mpc}$, and a prominent filamentary
structure associated with this galaxy cluster. A close look at
figure~\ref{fig:distribution} shows that we have significant deviation
from the ionization equilibrium state at the outskirts of galaxy
clusters and filamentary structures, where cosmological shocks occur
\citep{Kang2005}.

\section{EFFECT ON DETECTABILITY AND OBSERVABLES OF WHIM}

The deviations of ionization fractions of O\,{\sc vi}, O\,{\sc vii} and
O\,{\sc viii} from the ionization equilibrium state shown in the
previous section will affect the observational features of WHIM. In this
section, using the results obtained in the previous section, we present
to what extent the detectability and observables of WHIM through its
emission and absorption signatures are altered by relaxing the
assumption of the ionization equilibrium. For a mock observation of
WHIM, we construct a lightcone data up to a redshift of $z=0.3$ by
stacking 11 simulation cubes in the same way as described in
\citet{Yoshikawa2003}. The lightcone data has a $5^\circ \times
5^{\circ}$ field-of-view and contains the ionization fractions of ions
for each gas particle computed in the previous section as well as those
in the ionization equilibrium state.

\subsection{Detection through emission}

\begin{figure}[tbp]
 \leavevmode
 \begin{center}
  \FigureFile(120mm,120mm){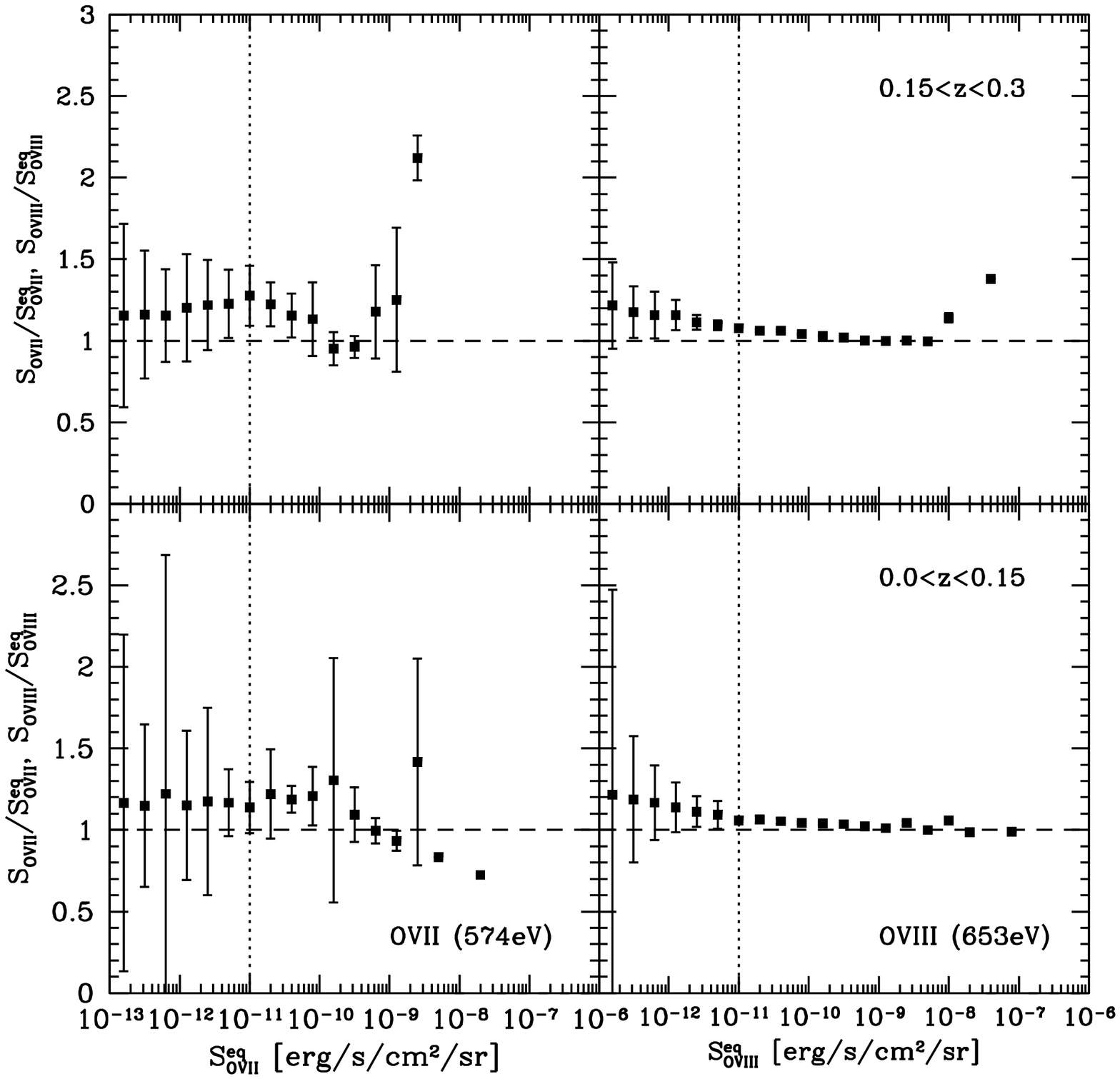}
 \end{center}
 \caption{Ratios of surface brightness of O\,{\sc vii} (574eV) and
 O\,{\sc viii} (653eV) emission lines for non-equilibrium states
 relative to equilibrium states as a function of surface brightness in
 equilibrium states. Upper and lower panels are for line emissions at
 redshift ranges of $0.15<z<0.3$ and $0.0<z<0.15$, respectively. Symbols
 and error bars indicate the averages and the standard deviations for
 each horizontal bin. Vertical dotted lines in both panels indicate the
 nominal detection limit of {\it DIOS}
 mission.\label{fig:emission_ratio}}
\end{figure}

\begin{figure}[htbp]
 \leavevmode
 \begin{center}
  \FigureFile(120mm,120mm){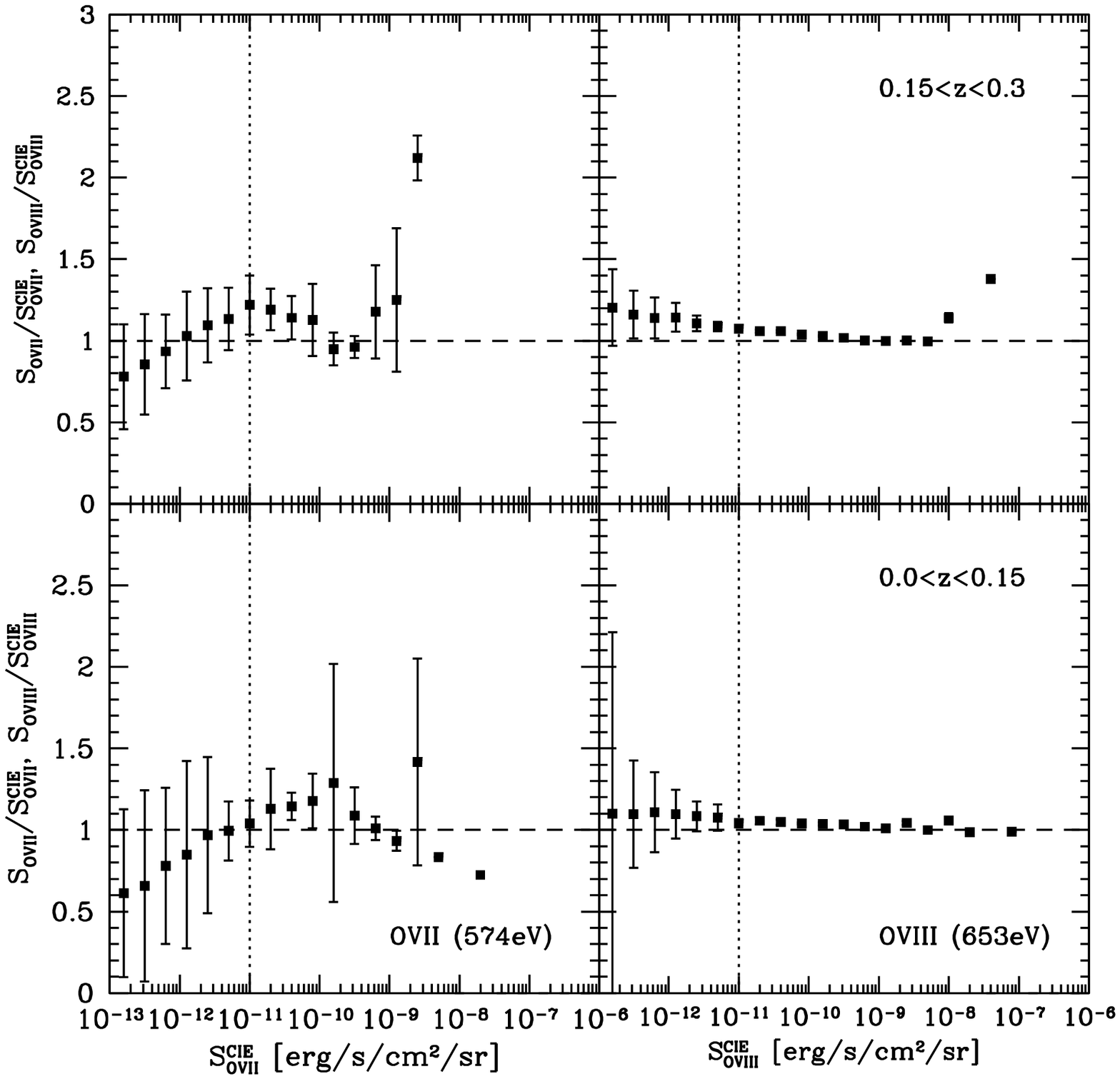}
 \end{center}
 \caption{Same as figure~\ref{fig:emission_ratio}, except that ratios
 relative to surface brightness in the collisional ionization
 equilibrium state are shown.\label{fig:emission_ratio_CIE}}
\end{figure}

Surface brightness of emission lines of O\,{\sc vii} at 574eV and
O\,{\sc viii} at 653eV ($S_{\rm OVII}$ and $S_{\rm OVIII}$,
respectively) are computed for both of the equilibrium and
non-equilibrium states. Here, we assume the latest specification of the
detector proposed for {\it DIOS} mission \citep{Ohashi2004}, namely a
spectroscopic energy resolution of 2 eV and a spatial resolution of 5
arcmin.  Figure~\ref{fig:emission_ratio} depicts the ratios of surface
brightness in the non-equilibrium state relative to that in the
ionization equilibrium state $S_{\rm OVII}/S^{\rm eq}_{\rm OVII}$ and
$S_{\rm OVIII}/S^{\rm eq}_{\rm OVIII}$, separately for line emitters at
redshift ranges of $0.0<z<0.15$ (lower panels) and $0.15<z<0.3$ (upper
panels). Vertical dotted lines in all the panels indicate the nominal
detection limit of {\it DIOS} mission ($10^{-11}$
erg\,s$^{-1}$\,cm$^{-2}$\,sr$^{-1}$). Note that we have significant
deviations of surface brightness ratios from unity at $\simeq 10^{-9}$
erg\,s$^{-1}$\,cm$^{-2}$\,sr$^{-1}$ in O\,{\sc vii} emission and $\simeq
10^{-8}$ erg\,s$^{-1}$\,cm$^{-2}$\,sr$^{-1}$ in O\,{\sc viii}
emission. It is found that these strong deviations take place at the
very central regions of galaxy clusters and are caused by shocks inside
them. One can see that surface brightness of O\,{\sc viii} (653 eV)
emission lines in the non-equilibrium ionization state is quite close to
that in the equilibrium state, and that O\,{\sc vii} (574 eV) emissions
are only slightly brighter than those in the ionization equilibrium
state above the detection limit of {\it DIOS} mission, irrespective of
redshift of the line emitters. Therefore, the detectability of WHIM
through its oxygen emission lines is not significantly different from
the theoretical predictions based on the assumption of ionization
equilibrium. Actually, the previous works on oxygen line emission of
WHIM by \citet{Yoshikawa2003}, \citet{Yoshikawa2004} and
\citet{Fang2005} all assume the collisional ionization
equilibrium. Therefore, it is useful to understand to what extent such
an assumption is acceptable. Figure~\ref{fig:emission_ratio_CIE} shows
the comparison of the surface brightness of O\,{\sc vii} and O\,{\sc
viii} emissions in the non-equilibrium state with those in the
collisional ionization equilibrium, and indicate that the assumption of
collisional ionization equilibrium is fairly valid for oxygen line
emitters brighter than the nominal detection limit of {\it DIOS},
$10^{-11}$ erg s$^{-1}$\,cm$^{-2}$\,sr$^{-1}$. Therefore, the numerical
predictions of detectability of WHIM through its oxygen line emission
presented in \citet{Yoshikawa2003} and \citet{Yoshikawa2004} is not
significantly altered by the effect of the non-equilibrium ionization
balance.

\subsection{Detection through absorption}

\begin{figure}[tbp]
 \leavevmode
 \begin{center}
  \FigureFile(120mm,120mm){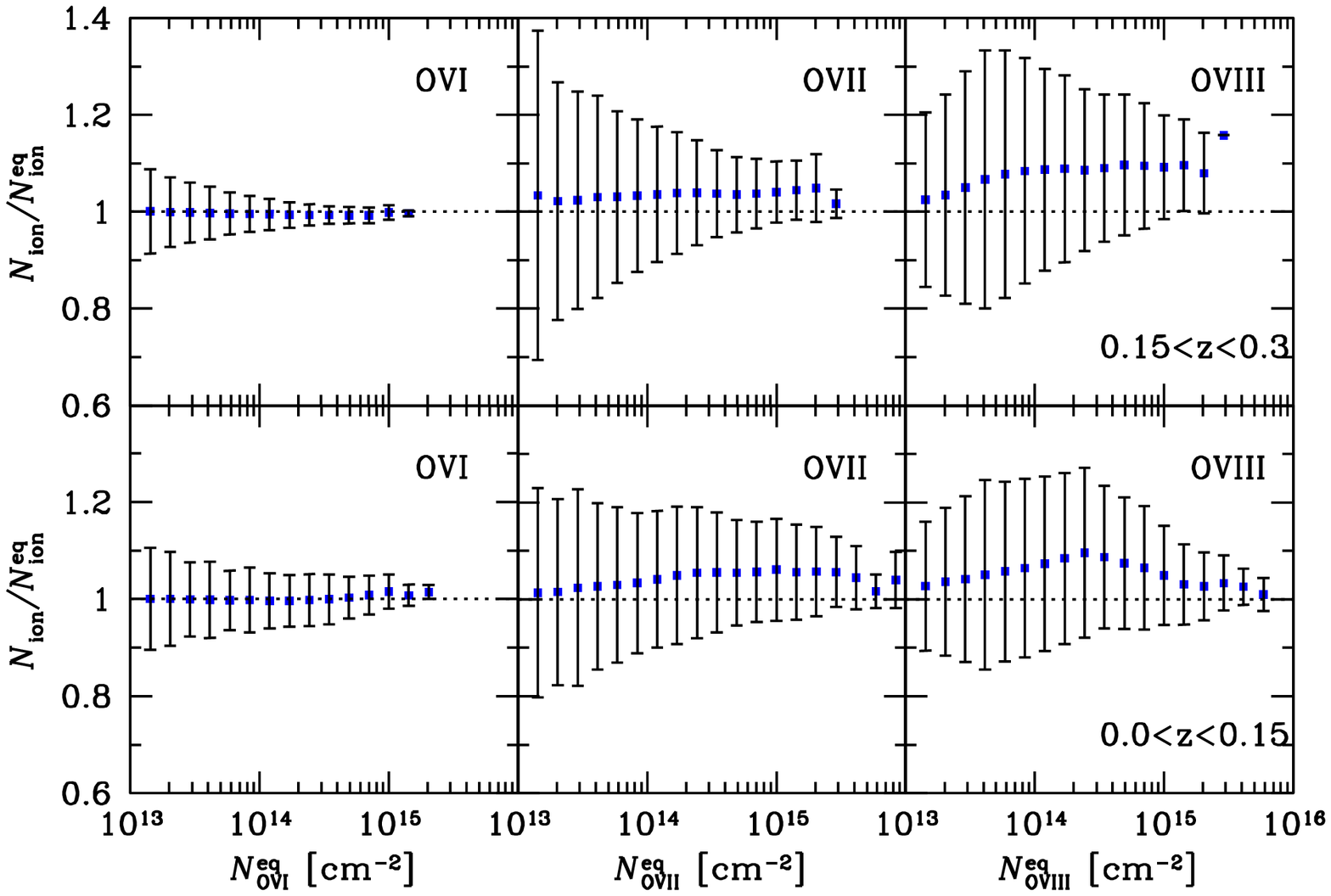}
 \end{center}
 \caption{Ratios of O\,{\sc vi}, O\,{\sc vii} and O\,{\sc viii} column
  densities of absorbers at redshift $0.0<z<0.15$ (lower panels) and
  $0.15<z<0.3$ (upper panels) in non-equilibrium states relative to
  those in ionization equilibrium states identified in the lightcone
  data as a function of column densities in ionization equilibrium
  states. Symbols and error bars indicate the averages and standard
  deviations for corresponding horizontal bins,
  respectively. \label{fig:column_density}}
\end{figure}

Column densities of O\,{\sc vi}, O\,{\sc vii} and O\,{\sc viii} ($N_{\rm
OVI}$, $N_{\rm OVII}$, and $N_{\rm OVIII}$, respectively) and those in
ionization equilibrium ($N^{\rm eq}_{\rm OVI}$, $N^{\rm eq}_{\rm OVII}$
and $N^{\rm eq}_{\rm OVIII}$) are calculated for $10^4$ randomly
selected line-of-sights in the $5^\circ \times 5^\circ$ field-of-view by
computing line integrals through the smoothing kernel of each gas
particle, and individual absorbers are identified in the same way as
done in \citet{Chen2003}. Each line-of-sight is divided into $\simeq
16000$ bins along redshift direction. Thus, each bin has a width of
$\simeq 5$ km s$^{-1}$ in velocity units.
Figure~\ref{fig:column_density} shows the column density ratios of the
identified absorbers in the non-equilibrium state relative to those in
the ionization equilibrium state ($N_{\rm OVI}/N^{\rm eq}_{\rm OVI}$,
$N_{\rm OVII}/N^{\rm eq}_{\rm OVII}$ and $N_{\rm OVIII}/N^{\rm eq}_{\rm
OVIII}$) as a function of the latter for absorbers at redshift ranges of
$0.0<z<0.15$ (lower panels) and $0.15<z<0.3$ (upper panels)
separately. As expected from the results shown in previous sections that
most of O\,{\sc vi} ions are quite close to the ionization equilibrium
state, column densities of O\,{\sc vi} absorbers are almost the same as
those in the ionization equilibrium state, irrespective of the redshift
of the absorbers. On the other hand, we have systematically larger
O\,{\sc vii} and O\,{\sc viii} column densities in the non-equilibrium
ionization state and also have large statistical dispersions. However,
the extent of difference in column densities between equilibrium and
non-equilibrium states is so mild that the effect on detectability of
WHIM through oxygen absorption lines is negligible.

\subsection{Line Ratios}

\begin{figure}[htbp]
 \leavevmode
 \begin{center}
  \FigureFile(120mm,120mm){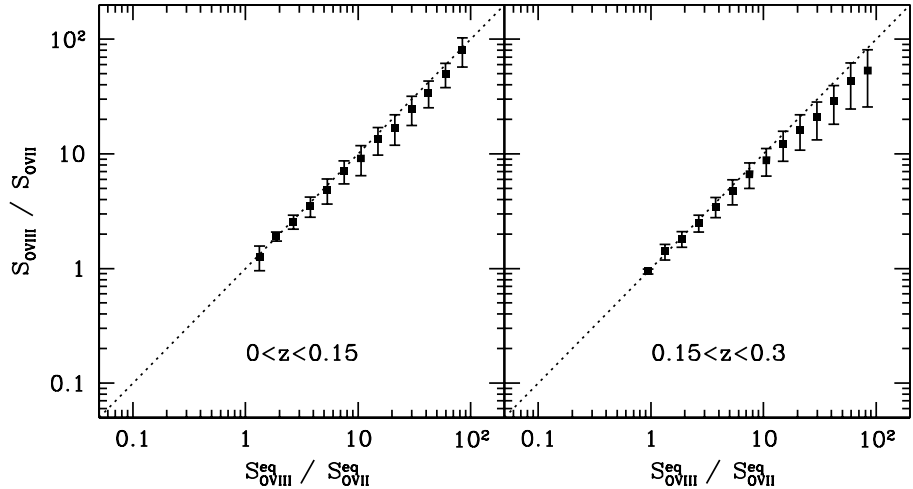}
 \end{center}
 \caption{Relations of line emission ratio $S_{\rm O\,VIII}/S_{\rm
 O\,VII}$ between non-equilibrium and equilibrium states.  Symbols and
 error bars indicate the averages and standard deviations for
 corresponding horizontal bins, respectively. \label{fig:emission_line_ratio}}
\end{figure}

\begin{figure}[htbp]
 \leavevmode
 \begin{center}
  \FigureFile(120mm,30mm){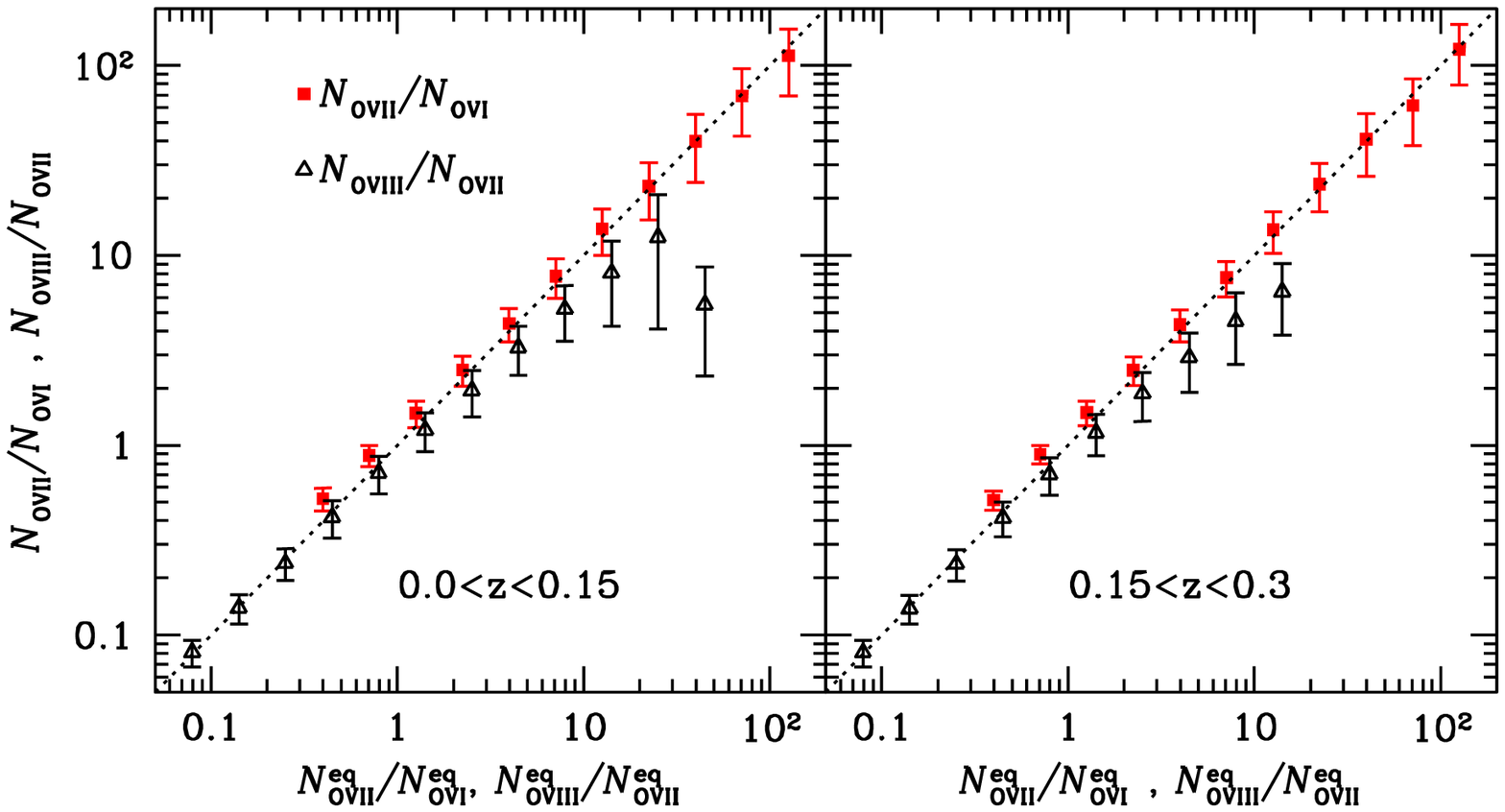}
 \end{center}
 \caption{Relations of column density ratios $N_{\rm O\,VII}/N_{\rm
 O\,VI}$ (filled squares) and $N_{\rm O\,VIII}/N_{\rm O\,VII}$ (open
 triangles) between non-equilibrium and equilibrium states. Symbols and
 error bars indicate the averages and standard deviations for
 corresponding horizontal bins, respectively.\label{fig:abs_line_ratio}}
\end{figure}

Apart from intensities of emission and absorption lines of WHIM, their
ratios are also a very important observables as diagnostics for the
nature of WHIM, in particular its temperature. Deviation from ionization
equilibrium will cause difference in ratios of emission and
absorption line intensities of WHIM from those with the assumption of
ionization equilibrium, and uncertainty in estimation of temperature of
WHIM.

Figure~\ref{fig:emission_line_ratio} depicts the relation of emission
line ratios $S_{\rm OVIII}/S_{\rm OVII}$ between equilibrium and
non-equilibrium states for line emitters at $0.0<z<0.15$ (left panels)
and $0.15<z<0.3$ (right panels). Here, line emitters with either $S^{\rm
eq}_{\rm OVII}$ or $S^{\rm eq}_{\rm OVIII}$ larger than $10^{-11}$
erg\,s$^{-1}$\,cm$^{-2}$\,sr$^{-1}$ are considered. One can see that the
line ratios $S_{\rm OVIII}/S_{\rm OVII}$ in the non-equilibrium state
are systematically lower than those in the ionization equilibrium state,
especially at $S_{\rm OVIII}/S_{\rm OVII}\gtrsim10$ for both redshift
ranges. Actually, we find that line emitters with $S_{\rm OVIII}$ or
$S_{\rm OVII} \simeq 10^{-11}$ erg\,s$^{-1}$\,cm$^{-2}$\,sr$^{-1}$
mainly contributes to these differences line ratios, and that line
ratios of emitters whose $S_{\rm OVII}$ and $S_{\rm OVIII}$ exceed
$10^{-10}$ erg\,s$^{-1}$\,cm$^{-2}$\,sr$^{-1}$ are statistically
consistent with those in the ionization equilibrium state.

Relation between column density ratios $N_{\rm OVII}/N_{\rm OVI}$ and
$N_{\rm OVIII}/N_{\rm OVII}$ in the ionization equilibrium and
non-equilibrium states are shown in figure~\ref{fig:abs_line_ratio} for
absorbers at redshift $0.0<z<0.15$ (left panel) and $0.15<z<0.3$ (right
panel), separately. Here, column density ratios $N_{\rm OVII}/N_{\rm
OVI}$ ($N_{\rm OVIII}/N_{\rm OVII}$) are considered for absorbers whose
$N_{\rm OVI}$ or $N_{\rm OVII}$ ($N_{\rm OVIII}$ or $N_{\rm OVII}$)
exceed 10$^{14}$ cm$^{-2}$. Although $N_{\rm OVII}/N_{\rm OVI}$ in the
non-equilibrium state statistically agrees well with that in the
ionization equilibrium state in both redshift ranges, the
non-equilibrium state exhibits systematically lower $N_{\rm
OVIII}/N_{\rm OVII}$ than the ionization equilibrium state. It is found
that these differences in $N_{\rm OVIII}/N_{\rm OVII}$ are due to
the absorbers with low ($\simeq 10^{14}$\,cm$^{-2}$) column density
of O\,{\sc vii} and O\,{\sc viii} ions. 

\begin{figure}[tbp]
 \leavevmode
 \begin{center}
  \rotatebox{270}{\FigureFile(80mm,80mm){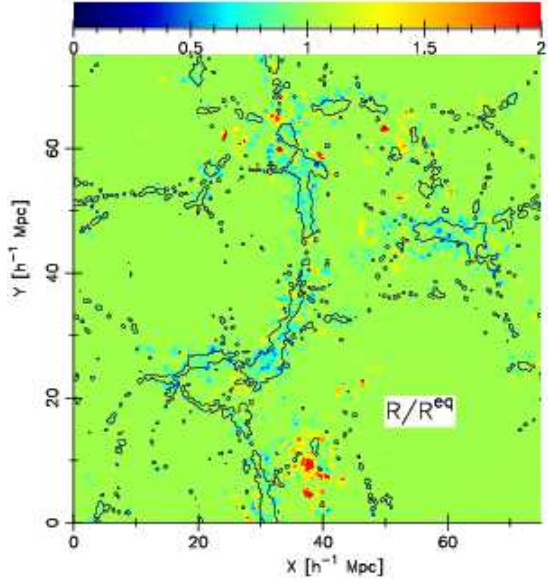}}
 \end{center}
 \caption{Spatial distribution of $R/R^{\rm eq}$ at $z=0$ in the same
 slice of the simulation volume as in
 figure~\ref{fig:distribution}. Contours for baryon mass density of
 $1+\delta_{\rm b}=1$ are also shown.\label{fig:ratio_distribution}}
\end{figure}

\begin{figure}[tbp]
 \leavevmode
 \begin{center}
  \rotatebox{270}{\FigureFile(60mm,60mm){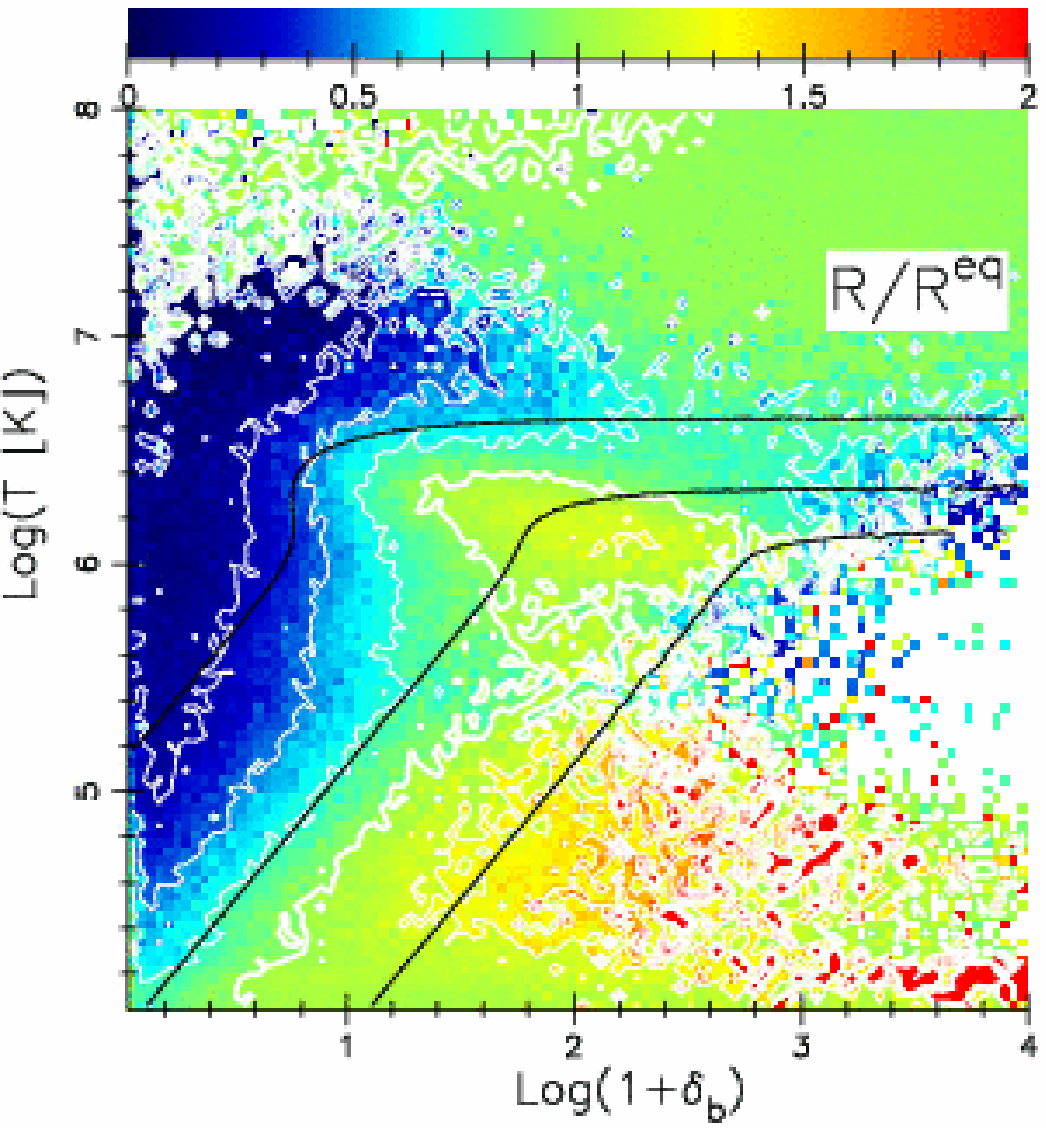}}
  \rotatebox{270}{\FigureFile(60mm,60mm){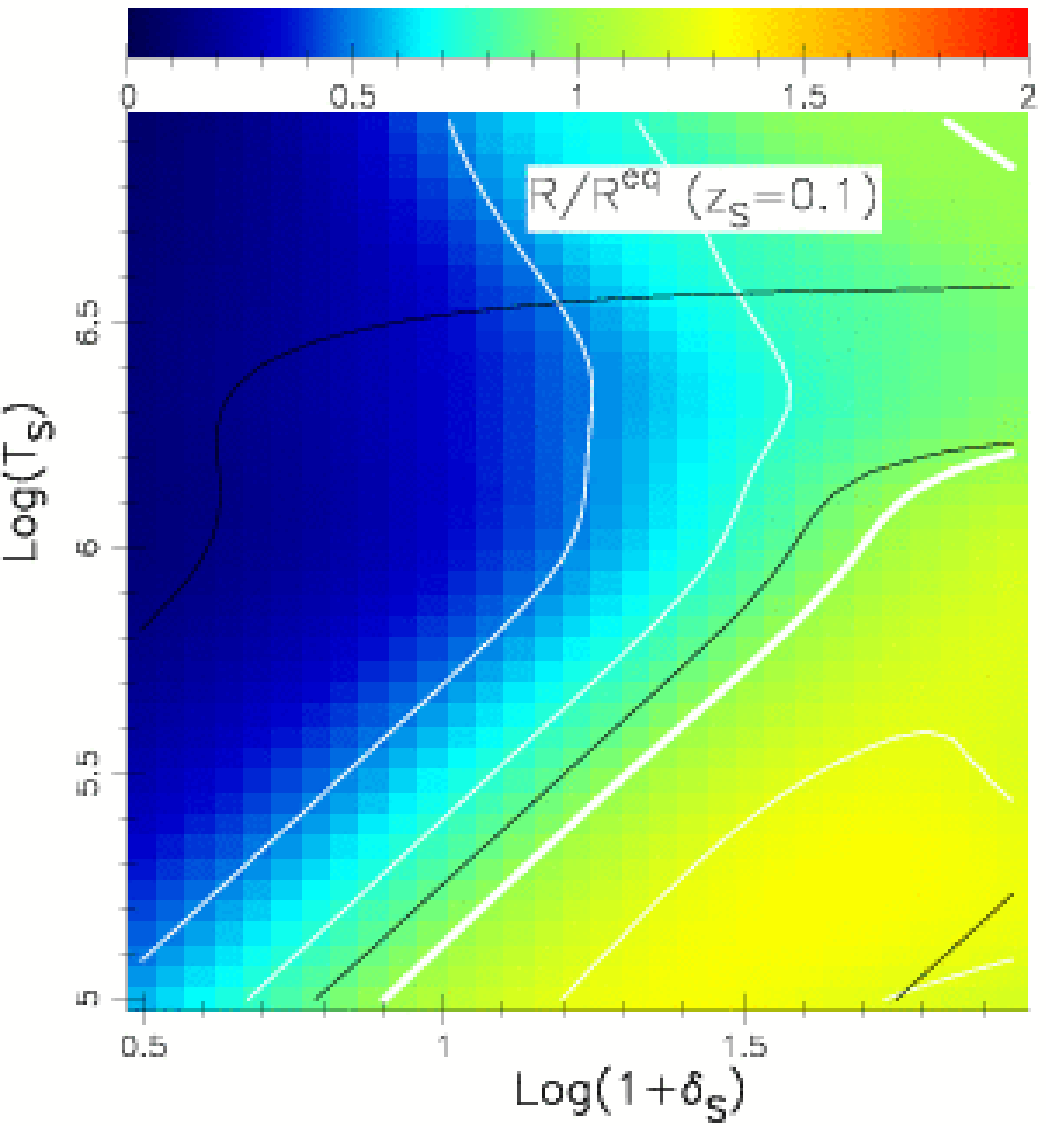}}
 \end{center}
 \caption{Maps of $R/R^{\rm eq}$ at $z=0$ on a $(1+\delta_{\rm b})$--$T$
 plane in a numerical simulation (left panel) and in a simple model with
 $z_{\rm s}=0.1$ (right panel). White contours are separated by a
 difference of 0.25 in $R/R^{\rm eq}$ and bold lines indicate $R/R^{\rm
 eq}=1$. Three black contours in each panel are for $R^{\rm eq}=0.1$, 1
 and 10 from bottom to top.\label{fig:ratio_contour_map}}
\end{figure}

The location and the physical conditions where the line ratio between
O\,{\sc viii} and O\,{\sc vii} deviates from that in ionization
equilibrium state are of great interest. Here we define the ratio of
ionization fractions of O\,{\sc viii} relative to O\,{\sc vii} in
equilibrium and non-equilibrium state as
\begin{equation}
 R\equiv\frac{f_{\rm OVIII}}{f_{\rm OVII}}
\end{equation}
and
\begin{equation}
 R^{\rm eq}\equiv\frac{f^{\rm eq}_{\rm OVIII}}{f^{\rm eq}_{\rm OVII}}.
\end{equation}
Note that $S_{\rm OVIII}/S_{\rm OVII}$ and $N_{\rm OVIII}/N_{\rm OVII}$
are roughly proportional to $R$.  Figure~\ref{fig:ratio_distribution}
shows the map of $R/R^{\rm eq}$ at a redshift of $z=0$ in the same slice
of the simulation volume as figure~\ref{fig:distribution}, and indicates
that we have $R/R^{\rm eq}\lesssim 1$ ubiquitously in filamentary
structures. This is why the line ratios, $S_{\rm OVIII}/S_{\rm OVII}$
and $N_{\rm OVIII}/N_{\rm OVII}$, are systematically smaller than those
in the ionization equilibrium state.  The left panel of
figure~\ref{fig:ratio_contour_map} shows the contour map of $R/R^{\rm
eq}$ on a $(1+\delta_{\rm b})$--$T$ plane, on which three contours for
$R^{\rm eq}=0.1$, 1 and 10 are overlayed from bottom to top.
Considering the fact that the $S_{\rm OVIII}/S_{\rm OVII}$ and $N_{\rm
OVIII}/N_{\rm OVII}$ in the non-equilibrium ionization state are lower
than their equilibrium values at $S^{\rm eq}_{\rm OVIII}/S^{\rm eq}_{\rm
OVII}\gtrsim 10$ and $N^{\rm eq}_{\rm OVIII}/N^{\rm eq}_{\rm
OVII}\gtrsim 10$, respectively, the left panel of
figure~\ref{fig:ratio_contour_map} indicates that the regions
responsible for such departures in line ratios from the equilibrium
state have a typical density of $1+\delta_{\rm b}\simeq 10$ and
temperature of $T\simeq10^{6-7}$K. The right panel of
figure~\ref{fig:ratio_contour_map} is the map of $R/R^{\rm eq}$ at $z=0$
in the simple model with $z_{\rm s}=0.1$ described in
section~\ref{sec:simple_model}, and shows the essentially same behavior
of $R/R^{\rm eq}$ on the $(1+\delta_{\rm b})$--$T$ plane. This suggests
that the regions with $R/R^{\rm eq}<1$ experienced a shock heating very
recently around a redshift of $z\simeq 0.1$, and that over abundant
O\,{\sc vii} ions due to the non-negligible timescale for the ionization
from O\,{\sc vii} to O\,{\sc viii} reduces $R/R^{\rm eq}$ to below
unity. Note that the loci with $R/R^{\rm eq}<1$ on the $(1+\delta_{\rm
b})$--$T$ plane in figure~\ref{fig:ratio_contour_map} and those with
$f_{\rm OVII}/f^{\rm eq}_{\rm OVII}>1$ roughly coincide (see upper right
panels of figure~\ref{fig:contour_map_simple_zs_0.1} and
\ref{fig:contour_map_simulation}).

\section{CONCLUSION AND DISCUSSION}

We investigate the non-equilibrium ionization state of oxygen ions in
WHIM by relaxing the assumption of ionization equilibrium and directly
integrating the time evolution of their ionization fractions along the
cosmological thermal histories of the baryons. 

First, we consider the general features of the non-equilibrium
ionization state for simple models of thermal histories of the cosmic
baryons, and find that ionization states of O\,{\sc vii}, O\,{\sc viii}
and O\,{\sc ix} ions are significantly different from those in the
ionization equilibrium state especially in typical physical conditions
for WHIM, while most of O\,{\sc vi} ion are close to the ionization
equilibrium state. These deviations from the ionization equilibrium are
caused by the fact that timescales for ionization and recombination
processes of these ions are comparable to or even longer than the hubble
time in low density plasma like WHIM. Furthermore, we find that rapid
decrease of the intensity of the UV and X-ray background radiation at
redshift $z<2$ also enlarge the deviation from the ionization
equilibrium state. This is because the timescale of the recombination
processes is so long that the recombination processes cannot catch up
with the ionization equilibrium state, in which ionization fractions
also vary with time as the UV and X-ray background radiation decays.

The evolution of ionization fractions of oxygen ions for thermal
histories obtained in a cosmological hydrodynamic simulation is also
investigated. The ionization state of oxygen ions at redshift of $z=0$
is similar to that obtained in the simple models of thermal histories in
which cosmological shocks take place at relatively recent epoch, say
$z\lesssim0.5$. Statistically, O\,{\sc vii} and O\,{\sc viii} ions are
over-abundant compared with the ionization equilibrium state in typical
physical conditions for WHIM, while O\,{\sc vi} ions are nearly in the
ionization equilibrium state. Spatial distribution of the oxygen ions
reveals that the deviations of ionization fractions of O\,{\sc vii} and
O\,{\sc viii} ions from the ionization equilibrium state take place at
the outskirts of galaxy clusters and the edge of filamentary structures.

We also investigate the effect of such deviations of oxygen ions from
the ionization equilibrium state on the detectability and observables of
WHIM through its emission and absorption line features. It is found that
although the surface brightness and column density of oxygen emission
and absorption lines slightly differ from those in the ionization
equilibrium state, the overall detectability of WHIM is almost the same
as what we have under the assumption of the ionization
equilibrium. However, the ratios of emission and absorption line
intensities between O\,{\sc vii} and O\,{\sc viii}, which are considered
to be important probes of temperature of WHIM, are significantly
different from those in the ionization equilibrium. This means that the
observed line ratios of WHIM may not reflect its real temperature. It is
found that the regions where the line ratios are different from those in
the ionization equilibrium are located at outer edges of filamentary
structures. Distribution of such regions on a density--temperature plane
is similar to the result of the simple model for the cosmological
thermal history with $z_{\rm s}=0.1$ described in
section~\ref{sec:simple_model}. These facts indicate that recent
cosmological shocks near the edge of filaments around a redshift of
$z\simeq 0.1$ cause the departure of line ratios from the ionization
equilibrium state.

In this paper, several important physical processes are missed. First of
all, in our simulation, the effects of energy feedback and metal
ejection from galaxies are ignored, and we assume that the metallicity
of the cosmic baryons is spatially uniform and constant irrespective of
redshift. Since oxygen atoms are mainly supplied by type-II supernovae
followed by the formation of massive stars, the spatial distribution of
oxygen is obviously time dependent and inhomogeneous. Furthermore, the
energy feedback by supernova explosions will affect the thermal history
of intergalactic medium (or WHIM) in the vicinity of
galaxies. Therefore, a self-consistent treatment of galaxy formation and
its feedback effect is required for better understanding of the
non-equilibrium ionization state of ions in WHIM. Recently,
\citet{Cen2006a} and \citet{Cen2006b} performed numerical simulations in
which the non-equilibrium ionization state of oxygen ions is considered
together with energy feedback and metal pollution caused by galaxy
formation. According to their results, energy feedback induces strong
shocks and additional non-equilibrium ionization state in the
intergalactic medium around galaxies if the energy feedback is strong
enough as performed in their simulations. Therefore, the deviation from
ionization equilibrium of oxygen ions in WHIM presented in this paper
can be regarded as its lower limit, and the actual deviations from the
ionization equilibrium may be much larger.

Secondly, it is assumed that equipartition between electrons and ions is
achieved instantaneously, or, in other words, electrons and ions always
have a locally common temperature. At cosmological shocks, by which most
of WHIM is heated, nearly all of the kinetic energy is converted into
the thermal energy of ions just after the shock. Actually, collisional
relaxation between ions and electrons through coulomb scattering takes a
long time which is comparable to the hubble time in a low density plasma
like WHIM. Therefore, as numerically shown by \citet{Takizawa1998} and
\citet{Yoshida2005}, the electron temperature $T_{\rm e}$ can be
systematically lower than the ion temperature $T_{\rm i}$ by a factor of
a few in the outskirts of galaxy clusters and WHIM, unless other
physical processes which convert the bulk kinetic energy into electron
thermal energy (e.g., \cite{Laming2000}) work efficiently. Since the
ionization and recombination rates $S_{i,j}$ and $\alpha_{j}$ in
equation~(\ref{eq:evolution}) depend on the electron temperature $T_{\rm
e}$ rather than the ion temperature $T_{\rm i}$, such difference between
$T_{\rm i}$ and $T_{\rm e}$ will affect the results obtained in this
paper. We will address this issue in near future by considering the
relaxation processes between ions and electrons.

\bigskip

We would like to thank an anonymous referee for his/her useful comments
and suggestions. Numerical computations presented in this paper were
carried out at the Astronomical Data Analysis Center of the National
Astronomical Observatory of Japan (project ID: wky17b). KY acknowledge
support from the Japan Society for the Promotion of Science.

\end{document}